\documentclass[prd, 10pt,nofootinbib, superscriptaddress, preprintnumbers,floatfix]{revtex4}

\usepackage{graphicx}
\usepackage{amssymb,amsfonts,amsmath}
\usepackage{hyperref}
\usepackage{nicefrac}
\usepackage{subfigure}
\usepackage[utf8]{inputenc}

\newcommand{\g}[1]{\gamma_{#1}}
\newcommand{\G}[1]{\gamma^{#1}}
\renewcommand{\l}{\left}
\renewcommand{\r}{\right}
\newcommand{\tr}{\mathrm{tr}}
\newcommand{\diag}{\mathrm{diag}}  
\newcommand{\bra}[1]{\left< #1 \right|} 
\newcommand{\ket}[1]{\left| #1 \right>} 
\newcommand{\chiral}[1]{\mathring{#1}} 
\newcommand{\kev}{\,\mathrm{keV}}
\newcommand{\mev}{\,\mathrm{MeV}}
\newcommand{\fm}{\,\mathrm{fm}}
\newcommand{\SU}[1]{\mathrm{SU}\l(#1\r)}

\newcommand{\experiment}{\mathrm{exp}}
\newcommand{\phys}{\mathrm{phys}}
\newcommand{\stat}{\mathrm{stat}}
\newcommand{\sys}{{\chi\mathrm{PT}}}
\newcommand{\tm}{\mathrm{tm}}
\newcommand{\etagg}{{\eta\gamma\gamma}}
\newcommand{\etapgg}{{\eta'\gamma\gamma}}
\newcommand{\etatogg}{{\eta\to\gamma\gamma}}
\newcommand{\etaptogg}{{\eta'\to\gamma\gamma}}
\newcommand{\hatFetagg}{\hat{F}_{\etagg^*}}
\newcommand{\hatFetapgg}{\hat{F}_{\etapgg^*}}
\newcommand{\limFetagg}[1]{\lim_{#1^2\to\infty} #1^2 F_{\etagg^*}(#1^2)}
\newcommand{\limFetapgg}[1]{\lim_{#1^2\to\infty} #1^2 F_{\etapgg^*}(#1^2)}

\newcommand{\Getatogg}{\Gamma_{\etatogg}}
\newcommand{\Getaptogg}{\Gamma_{\etaptogg}}

\newcommand{\Mpi}{M_\pi}
\newcommand{\MK}{M_\mathrm{K}}
\newcommand{\Meta}{M_\eta}
\newcommand{\Metap}{M_{\eta^\prime}}

\newcommand{\chibar}{\overline{\chi}}

\begin{document}

\title{Flavor-singlet meson decay constants from $N_f=2+1+1$ twisted mass lattice QCD}

\newcommand\bn{Helmholtz-Institut f\"ur Strahlen- und Kernphysik and Bethe Center for Theoretical Physics, University of Bonn, D-53115 Bonn, Germany}
\newcommand\mz{Institut f\"ur Kernphysik, University of Mainz, D-55128 Mainz, Germany}
\author{Konstantin Ottnad}\email{ottnad@hiskp.uni-bonn.de}\affiliation{\bn}\affiliation{\mz}
\author{Carsten Urbach}\affiliation{\bn}
\collaboration{ETM Collaboration}\noaffiliation

\begin{abstract}
 \vspace*{0.5cm}
 \begin{center}
  \includegraphics[draft=false,width=.18\linewidth]{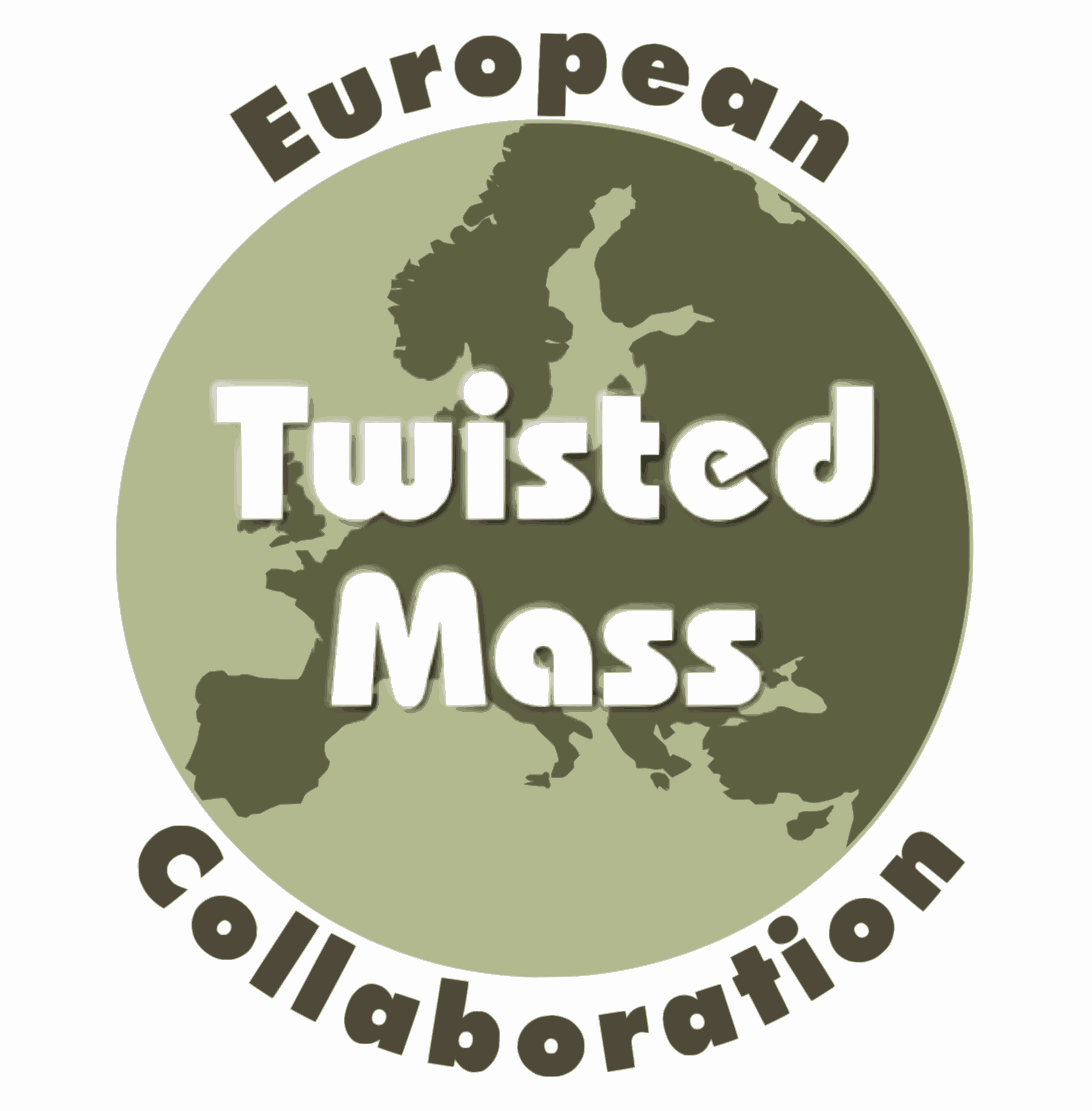}
 \end{center}
 \vspace*{0.5cm}
 We present an improved analysis of our lattice data for the $\eta$--$\eta'$ system, including a correction of the relevant correlation functions for residual topological finite size effects and employing consistent chiral and continuum fits. From this analysis we update our physical results for the masses $M_\eta=557(11)_\stat(03)_\sys\mev$ and $M_\eta'=911(64)_\stat(03)_\sys\mev$, as well as the mixing angle in the quark flavor basis $\phi=38.8(2.2)_\stat(2.4)_\sys^\circ$ in excellent agreement with other results from phenomenology. Similarly, we include an analysis for the decay constant parameters, leading to $f_l=125(5)_\stat(6)_\sys\mev$ and $f_s=178(4)_\stat(1)_\sys\mev$. The second error reflects the uncertainty related to the chiral extrapolation. The data used for this study has been generated on gauge ensembles provided by the European Twisted Mass Collaboration with $N_f=2+1+1$ dynamical flavors of Wilson twisted mass fermions. These ensembles cover a range of pion masses from $220\mev$ to  $500\mev$ and three values of the lattice spacing. Combining our data with a prediction from chiral perturbation theory, we give an estimate for the physical $\eta,\eta' \rightarrow \gamma\gamma$ decay widths and the singly-virtual $\eta,\eta'\rightarrow\gamma\gamma^*$ transition form factors in the limit of large momentum transfer.
\end{abstract}

\maketitle
\clearpage

\section{Introduction} 
\label{sec:intro}

The axial anomaly and the topological nature of quantum chromodynamics (QCD) are
able to explain the large experimentally observed mass of the
$\eta^\prime$ meson. This understanding was possible due to
perturbative arguments leading to the Witten-Veneziano~\cite{Witten:1979vv,Veneziano:1979ec}
formula. This was recently confirmed nonperturbatively using lattice
simulations in Ref.~\cite{Cichy:2015jra}. 

Beyond masses, there is also
the phenomenon of mixing: $\eta$ and $\eta^\prime$ mesons are not
flavor eigenstates, but represent a mixing of an octet and a
singlet state. In contrast to the $\omega$--$\phi$ meson mixing in the
vector channel, the mixing in the pseudoscalar channel is ideal at the
SU$(3)$ symmetric point with mixing angle $\phi=54.7^\circ$ (in the
quark flavor basis), but exhibits significantly smaller $\phi$-values at
physical quark masses~\cite{Michael:2013gka}. The precise knowledge of
the mixing angle is important for several phenomenological
applications, most notably for improving the theoretical estimate of
the hadronic contribution to the anomalous magnetic moment of the
muon~\cite{Escribano:2013kba}. 

The investigation of $\eta$ and $\eta^\prime$ mesons requires a
nonperturbative method, provided by lattice QCD. However, such an
investigation is challenging due to large contributions by so-called
fermionic disconnected diagrams. Flavor-singlet pseudoscalar
mesons have been studied using lattice QCD before. For $N_f=2$ results
can be found in
Refs.~\cite{Lesk:2002gd,Allton:2004qq,Hashimoto:2008xg,Jansen:2008wv,Helmes:2017ccf}. For
$N_f=2+1$ or $N_f=2+1+1$ they were 
studied in
Refs.~\cite{Kaneko:2009za,Christ:2010dd,Dudek:2011tt,Gregory:2011sg,Dudek:2013yja,Ottnad:2012fv,Michael:2013gka,Michael:2013vba,Bali:2014pva,Bali:2017qce}. For 
an approach based on a purely gluonic operator see Ref.~\cite{Fukaya:2015ara}.

In this paper we extend our
previous studies~\cite{Ottnad:2012fv,Michael:2013gka,Michael:2013vba} of properties of
$\eta$ and $\eta^\prime$ mesons in two ways: 
first by an improved analysis: on an enlarged number of Monte Carlo
ensembles we perform consistent chiral and continuum
extrapolations. This leads to slightly changed results when compared
to Ref.~\cite{Michael:2013gka}, mostly within the quoted error
bars, with the exception being the mixing angle $\phi$. The larger
change in $\phi$ comes from the fact that we are now able to resolve
the lattice spacing dependence in $\phi$, too, due to more ensembles
at the finest lattice spacing value. Moreover, by using derivatives of
correlation functions instead of the correlation function themselves,
we are able to remove systematic effects from not optimally sampled
topological sectors. 

Second, for the first time we estimate flavor-singlet pseudoscalar
decay constants from lattice QCD. We use these to
determine also physical $\eta,\eta' \rightarrow \gamma\gamma$ decay
widths and the singly-virtual $\eta,\eta'\rightarrow\gamma\gamma^*$ 
transition form factors in the limit of large momentum transfer. A
first account of this work can be found in Ref.~\cite{Michael:2013vba}.

For the determinations of the aforementioned decay constants we rely
on chiral perturbation theory instead of determining them directly
from flavor-singlet axial-vector matrix elements. The reason for
this procedure is an unfavorable signal-to-noise ratio in some of
those matrix elements, which prevents a meaningful analysis. 
The results are compared to phenomenological determinations mostly
summarized in Ref.~\cite{Feldmann:1999uf}.

The results presented here are based on gauge configurations produced
by the European Twisted Mass Collaboration (ETMC) with active up/down,
strange and charm quarks. Based on three values of the lattice spacing
and pion masses ranging from $220$ to $500$ MeV, the ETMC ensembles
allow us to perform a controlled chiral and continuum
extrapolation. Dedicated ensembles with varied strange quark masses
let us control also the strange quark mass dependence.

\section{Lattice setup} \label{sec:setup}
The calculations for this work have been performed on gauge configurations generated by the European Twisted Mass Collaboration (ETMC) \cite{Baron:2010bv,Baron:2010th,Baron:2011sf} using $N_f=2+1+1$ dynamical quark flavors of twisted Wilson fermions at maximal twist. In the gauge sector the Iwasaki action \cite{Iwasaki:1985we,Iwasaki:1996sn} has been used in the generation of configurations
\begin{equation}
 S_G[U]=\frac{\beta}{3}\sum_x\left(b_0\sum_{\genfrac{}{}{0pt}{}{\mu,\nu=1}{1\leq\mu<\nu}}^4\mathrm{Re}\,\tr \left(1-P^{1\times 1}_{x;\mu\nu}\right)+b_1\sum_{\genfrac{}{}{0pt}{}{\mu,\nu=1}{\mu \neq \nu}}^4\mathrm{Re}\, \tr \left(1-P^{1\times 2}_{x;\mu\nu}\right)\right)\,,
 \label{eq:iwasaki_action}
\end{equation}
where $b_1=-0.331$ and $b_0=1-8b_1$. $P^{1\times 1}_{x;\mu\nu}$ and $P^{1\times 2}_{x;\mu\nu}$ denote quadratic (plaquette) and rectangular Wilson loops composed of gauge links. \par

In the twisted basis the fermionic action containing a mass-degenerate, light quark doublet $\chi_l=(\chi_u,\chi_d)^T$ reads \cite{Frezzotti:2000nk,Frezzotti:2003ni,Frezzotti:2004wz}
\begin{equation}
  S_l[\chi_l,\chibar_l,U]=a^4 \sum_x \chibar_l(x)(D_W[U]+m_0+i\mu_l\gamma_5\tau_3)\chi_l(x)\,,
 \label{eq:light_action}
\end{equation}
while for a nondegenerate, heavy quark doublet $\chi_h=(\chi_c,\chi_s)^T$ we have \cite{Frezzotti:2004wz,Frezzotti:2003xj}
\begin{equation}
S_h[\chi_h,\chibar_h,U]=a^4 \sum_x\chibar_h(x)(D_W[U]+m_0+i\mu_\sigma\gamma_5\tau_1+\mu_\delta\tau_3)\chi_h(x)\,,
 \label{eq:heavy_action}
\end{equation}
where the Pauli matrices $\tau^i$, $i=1,2,3$ act in flavor space. The massless Wilson Dirac operator
\begin{equation}
 D_W = D_W=\frac{1}{2}(\gamma_\mu(\nabla_\mu+\nabla^\star)-a\nabla^\star_\mu\nabla_\mu)\,,
 \label{eq:D_W}
\end{equation}
depends implicitly on the gauge links $U$. The doublets $\chi_l$ and $\chi_h$ are related to doublet fields in the physical basis $\psi_l$, $\psi_h$ via chiral rotations. The bare strange and charm quark masses $m_s$, $m_c$ are given in terms of the bare input parameters $\mu_\delta$ and $\mu_\sigma$
\begin{equation}
 \mu_{c,s}=\mu_\sigma\pm Z\mu_\delta \,,
 \label{eq:heavy_quark_masses}
\end{equation}
where $Z=Z_P/Z_S$ denotes the ratio of pseudoscalar and scalar flavor nonsinglet renormalization factors. The renormalized quark masses require an additional factor of nonsinglet $1/Z_P$
\begin{equation}
 \mu_{c,s}^r = Z_P^{-1} \mu_\sigma\pm Z_S^{-1} \mu_\delta \,,
 \label{eq:heavy_quark_masses_renormalized}
\end{equation}
which is the same as for the light bare quark mass, i.e. $\mu_l^r= \mu_l / Z_P$. \par

\begin{table}[t!]
 \centering
 \begin{tabular*}{.8\textwidth}{@{\extracolsep{\fill}}lllllllrrr}
  \hline\hline
  ensemble  & $\beta$ & T/a & L/a & $a\mu_l$ & $a\mu_\sigma$ & $a\mu_\delta$ & $N_\mathrm{conf}$ & $\Delta N$ & $N_S$ \\
  \hline\hline
  A30.32   & 1.90 & 64 & 32 & 0.0030 & 0.150  & 0.190  & 1363 &  4 &  24 \\
  A40.32   & 1.90 & 64 & 32 & 0.0040 & 0.150  & 0.190  &  863 &  4 &  24 \\
  A40.24   & 1.90 & 48 & 24 & 0.0040 & 0.150  & 0.190  & 1877 &  4 &  32 \\
  A60.24   & 1.90 & 48 & 24 & 0.0060 & 0.150  & 0.190  & 1248 &  4 & 128 \\
  A80.24   & 1.90 & 48 & 24 & 0.0080 & 0.150  & 0.190  & 2449 &  2 &  32 \\
  A100.24  & 1.90 & 48 & 24 & 0.0100 & 0.150  & 0.190  & 2489 &  2 &  32 \\
  \hline
  A80.24s  & 1.90 & 48 & 24 & 0.0080 & 0.150  & 0.197  & 2514 &  2 &  32 \\
  A100.24s & 1.90 & 48 & 24 & 0.0100 & 0.150  & 0.197  & 2312 &  2 &  32 \\
  \hline 
  B25.32   & 1.95 & 64 & 32 & 0.0025 & 0.135  & 0.170  & 1467 &  4 &  24 \\
  B35.32   & 1.95 & 64 & 32 & 0.0035 & 0.135  & 0.170  & 1251 &  4 &  24 \\
  B55.32   & 1.95 & 64 & 32 & 0.0055 & 0.135  & 0.170  & 4996 &  4 &  48 \\
  B75.32   & 1.95 & 64 & 32 & 0.0075 & 0.135  & 0.170  &  922 &  8 &  24 \\
  B85.24   & 1.95 & 48 & 24 & 0.0085 & 0.135  & 0.170  &  573 & 10 &  32 \\
  \hline
  D15.48   & 2.10 & 96 & 48 & 0.0015 & 0.120  & 0.1385 & 1034 &  2 &  24 \\
  D20.48   & 2.10 & 96 & 48 & 0.0020 & 0.120  & 0.1385 &  429 &  4 &  24 \\
  D30.48   & 2.10 & 96 & 48 & 0.0030 & 0.120  & 0.1385 &  458 &  8 &  24 \\
  D45.32sc & 2.10 & 64 & 32 & 0.0045 & 0.0937 & 0.1077 & 1074 &  4 &  48 \\
  \hline\hline
  \vspace*{0.1cm}
 \end{tabular*}
 \caption{Overview of ensembles and respective input parameters. In addition, we give the total number of gauge configurations $N_\mathrm{conf}$ used, the spacing in terms of HMC trajectories between two adjacent configurations used in our study $\Delta N$ and the number of stochastic samples for the computation of the quark disconnected diagrams $N_S$. The spatial extend $L/a$ satisfies $T/a = 2 \cdot L/a$ on all ensembles.}
 \label{tab:ensembles}
\end{table}

We employ $17$ gauge ensembles as detailed in Table~\ref{tab:ensembles} at three different values of $\beta$ corresponding to three different values of the lattice spacing $a$, cf. Table~\ref{tab:beta_r0_a_Z}. Compared to previous studies of the $\eta$,$\eta'$--system \cite{Michael:2013gka,Cichy:2015jra}, we have added one more ensemble at the finest lattice spacing (D20.48) and significantly increased the statistic on the B55.32 ensemble. In general, all observables have been computed with the full statistic as given in Table~\ref{tab:ensembles} with exception of the kaon mass $M_K$ and the kaon decay constant $f_K$, that have been computed only on a subset of configurations in many cases. However, this is sufficient for our purposes as we are neither interested in $M_K$ nor $f_K$ themselves in this study. The resulting errors for derived observables are always dominated by the statistical uncertainties in the flavor-singlet sector, anyways. \par

\begin{table}[t!]
 \centering
 \begin{tabular*}{.8\textwidth}{@{\extracolsep{\fill}}lllll}
  \hline\hline
  $\beta$  & $r_0/a$ & $a\,[\mathrm{fm}]$ & Z (M1) & Z (M2) \\
  \hline\hline
  1.90 & 5.31(8) & 0.0885(36) & 0.699(13) & 0.651(06) \\
  1.95 & 5.77(6) & 0.0815(30) & 0.697(07) & 0.666(04) \\
  2.10 & 7.60(8) & 0.0619(18) & 0.740(05) & 0.727(03) \\
  \hline\hline
  \vspace*{0.1cm}
 \end{tabular*}
 \caption{Values of $r0/a$, $a$ and $Z$ corresponding to the three values of $\beta$ \cite{Carrasco:2014cwa}. The labels M1 and M2 refer to two methods used to determine renormalization constants in this reference.}
 \label{tab:beta_r0_a_Z}
\end{table}

Table~\ref{tab:beta_r0_a_Z} also contains the results for $Z$ at each value of $\beta$, which are needed for the computation of matrix elements. The labels M1 and M2 refer to the two different methods used in Ref.~\cite{Carrasco:2014cwa} for the determination of renormalization factors. In addition, we included the values for the Sommer scale $r_0$ at each value of the lattice spacing, $r_0/a$, that were taken again from Ref.~\cite{Carrasco:2014cwa} together with the physical value
\begin{equation}
 r_0 = 0.474(14)_\stat \fm \,,
 \label{eq:r_0}
\end{equation}
which is required to set the scale in our study. Note that in an earlier publication in Ref.~\cite{Michael:2013gka} slightly different values have been used for $r_0$ and $r_0/a$. However, those are now superseded by the values from the final analysis in Ref.~\cite{Carrasco:2014cwa} that we use here. \par

While the values of $a\mu_\sigma$ and $a\mu_\delta$ defining the bare strange and charm masses are generally fixed for each choice of $\beta$, we include a few dedicated ensembles (A80.24s, A100.24.s and D45.32sc), which have been generated with different $\mu_{\delta}$ -- and in case of D45.32sc -- also different $\mu_{\sigma}$ values. This allows us to explicitly resolve the dependence on the strange quark mass and obtain more stable results from chiral fits. \par

For the computation of quark-disconnected diagrams we employ stochastic volume sources. The corresponding number of stochastic samples $N_S$ is included in Table~\ref{tab:ensembles} and is chosen such that the final statistical errors are dominated by gauge noise. The statistical errors for all observables are computed using the blocked bootstrap with 10000 samples and blocklengths chosen such that the effective length of every block corresponds to at least 20 HMC trajectories. This has been found sufficient to deal with autocorrelations in an earlier study in Ref.~\cite{Ottnad:2012fv}. \par

\section{Computation of masses and amplitudes}
The extraction of masses and matrix elements for the $\eta$,$\eta'$--system has already been discussed in detail in previous publications \cite{Ottnad:2012fv,Michael:2013gka,Michael:2013vba,Cichy:2015jra}. In the following we will first briefly summarize the relevant methods and then introduce a modification leading to systematic improvement of our existing analysis. This improvement concerns a possible contamination of the large--$t$ behavior of the flavor-singlet correlations functions induced by imperfectly sampled topology. Finally, we detail the extraction of mixing parameters and further, derived observables such as decay widths. \par

\subsection{Correlation function matrix}
In the physical basis we consider the following three local pseudoscalar operators
\begin{align}
 \mathcal{P}_l^{0,\phys}(x) &= \frac{1}{\sqrt{2}} \bar{\psi}_l(x) i \g{5} \psi_l(x) \,, \label{eq:light_op_phys}\\
 \mathcal{P}_h^{\pm,\phys}(x) &= \bar{\psi}_h(x) i \g{5} \frac{1\pm\tau^3}{2}\psi_h(x) \,, \label{eq:heavy_op_phys}
\end{align}
where $\bar{\psi}_l(x)$, $\psi_l(x)$ and $\bar{\psi}_h(x)$, $\psi_h(x)$ denote degenerate light and nondegenerate heavy quark doublets, respectively. While the doublet structure of the fields is required for the rotation to the twisted mass basis, the flavor projector $(1\pm\tau^3)/2$ disentangles charm (``$+$'') and strange (``$-$'') contributions. Upon rotation to the twisted basis the operators read at maximal twist
\begin{align}
 \mathcal{P}_l^{0,\phys}(x) &\rightarrow -\frac{1}{\sqrt{2}}\bar{\chi}_l(x) \tau^3 \chi_l(x) \equiv \mathcal{S}_l^{3,\tm}(x) \,, \label{eq:light_S_tm_eta_operator}\\
 \mathcal{P}_h^{\pm,\phys}(x) &\rightarrow \frac{1}{2}\bar{\chi}_h(x) \l(-\tau^1 \pm i \g{5} \tau^3 \r) \chi_h(x) \equiv \mathcal{P}_h^{\pm,\tm}(x) \,. \label{eq:heavy_P_tm_eta_operators}
\end{align}
In Refs.~\cite{Ottnad:2012fv,Michael:2013gka} it has been shown that the charm quark operator essentially has no overlap with the $\eta$,$\eta'$--states and can hence be neglected. Therefore, we drop it from the actual analysis and keep only light and strange quark operators, i.e. $\mathcal{S}_l^{\tm}(x) \equiv \mathcal{S}_l^{3,\tm}(x)$ and $\mathcal{P}_s^{\tm}(x) \equiv \mathcal{P}_h^{-,\tm}(x) $. Considering renormalization, the operators can be written as
\begin{align}
 \mathcal{S}_l^{\tm,r}(x) &= \frac{1}{\sqrt{2}} Z_S \bar{\chi}_l(x) \tau^3 \chi_l(x) \,, \label{eq:light_op_renormalized} \\
 \mathcal{P}_s^{\tm,r}(x) &= \frac{1}{2} Z_P \bar{\chi}_h(x) \l(-\frac{Z_S}{Z_P}\tau^1 - i \g{5} \tau^3\r) \chi_h(x) \,. \label{eq:strange_op_renormalized}
\end{align}
Pulling out a factor $Z_P^2$ from the resulting $2\times2$ correlation function matrix we have
\begin{equation}
 \mathcal{C}^r(t) = Z_P^2 \tilde{\mathcal{C}}(t) \,,
\end{equation}
where
\begin{equation}
 \tilde{\mathcal{C}}(t) = \l(\begin{array}{cc}
 \bigl<\tilde{\mathcal{S}}_l^{\tm}(t)\tilde{\mathcal{S}}_l^{\tm}(0)\bigr> &
 \bigl<\tilde{\mathcal{S}}_l^{\tm}(t)\tilde{\mathcal{P}}_s^{\tm}(0)\bigr> \\
 \bigl<\tilde{\mathcal{P}}_s^{\tm}(t)\tilde{\mathcal{S}}_l^{\tm}(0)\bigr> &
 \bigl<\tilde{\mathcal{P}}_s^{\tm}(t)\tilde{\mathcal{P}}_s^{\tm}(0)\bigr> \end{array}\r)
 \label{eq:corr_matrix}
\end{equation}
contains operators that are projected to zero-momentum and renormalized up to a global factor of flavor nonsinglet $Z_P$, i.e.
\begin{align}
 \tilde{\mathcal{S}}_l^{\tm}(x) &= \frac{1}{\sqrt{2}} Z^{-1} \bar{\chi}_l(x) \tau^3 \chi_l(x)  \,, \label{eq:light_op_Z_only} \\
 \tilde{\mathcal{P}}_s^{\tm}(x) &= \frac{1}{2} \bar{\chi}_h(x) \l(-Z^{-1}\tau^1 - i \g{5} \tau^3\r) \chi_h(x) \,. \label{eq:heavy_op_Z_only}
\end{align} 
The mixing of flavor nonsinglet pseudoscalar and scalar currents in the heavy quark basis remains manifest in the corresponding ratio of renormalization constants $Z$. Note that the construction of these operators requires only the ratio $Z$ instead of $Z_P$ and $Z_S$ separately. We will show later that renormalization of the correlation function matrix up to a global factor $Z_P^2$ is sufficient for the calculation of the mixing parameters, as this factor will be absorbed by the renormalization of corresponding factors of quark masses. \par

As first proposed in Ref.~\cite{Neff:2001zr} and subsequently used in Refs.~\cite{Jansen:2008wv,Michael:2013gka}, we replace the quark-connected pieces in the correlation functions by the ground state contribution. This allows to extract the $\eta$ and $\eta'$ states from the resulting principal correlators from the earliest available timeslice $t_1^{\eta,\eta'}=2a$ on, leading to a significant improvement in the signal-to-noise ratio. For further technical details on this procedure, including fit parameters and results, we refer to Appendix~\ref{sec:appendix_A} and tables therein. \par

\subsection{Correlator improvement for topological effects}
\begin{figure}[t]
 \centering
 \subfigure[]{\includegraphics[width=.48\linewidth]{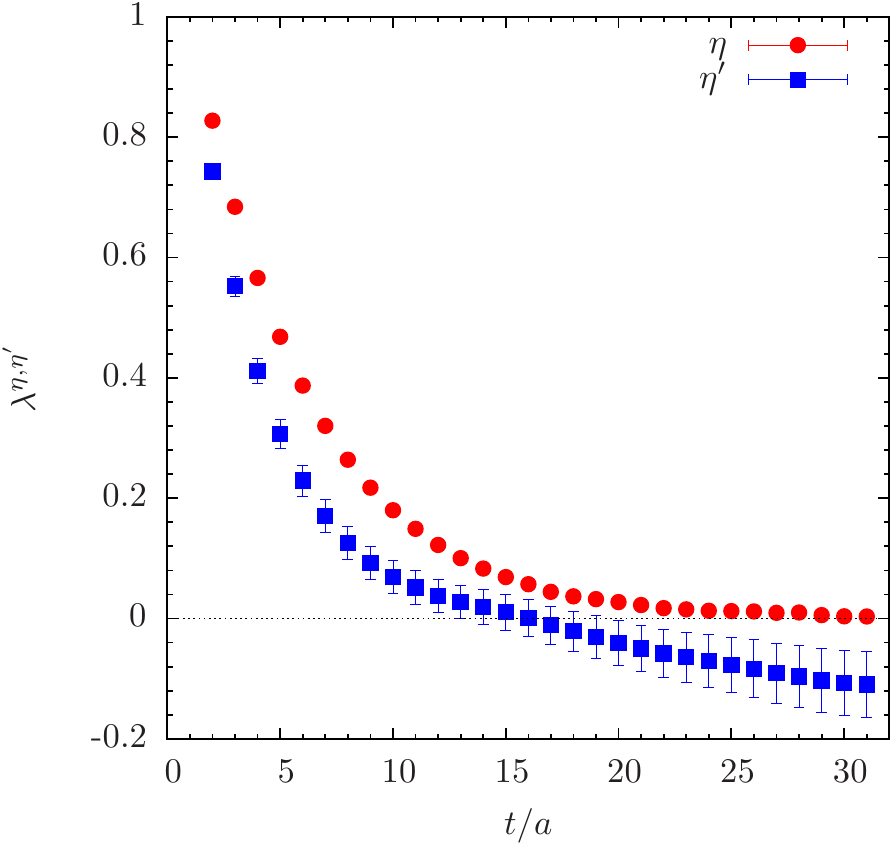}}\quad
 \subfigure[]{\includegraphics[width=.48\linewidth]{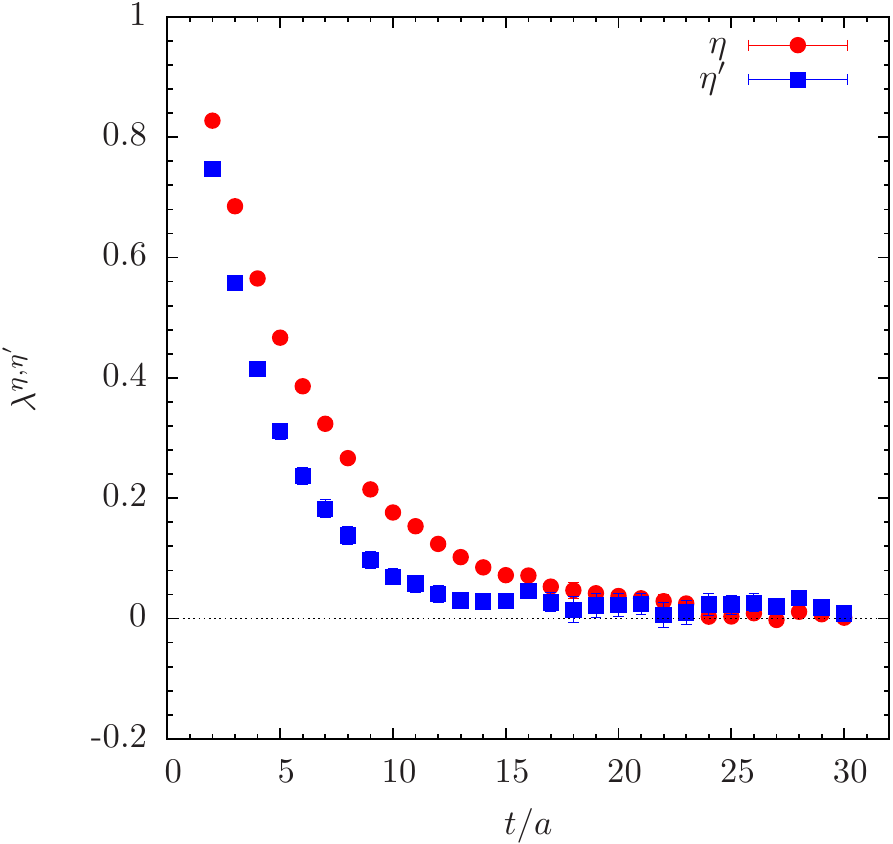}}\quad
 \caption{(a) Eigenvalues $\lambda^{\eta,\eta'}$ from solving the GEVP Eq.~(\ref{eq:GEVP}) for the correlation function matrix in Eq.~(\ref{eq:corr_matrix}). (b) Same as left panel, but from solving the GEVP for the derivative correlation function matrix defined in Eq.~(\ref{eq:shifted_corr_matrix})}
 \label{fig:eigenvalues}
\end{figure}

In Ref.~\cite{Aoki:2007ka} it has been pointed out that the large--$t$ behavior of quark-disconnected contribution $C_\mathrm{2pt}^\mathrm{disc}(t)$ to pseudoscalar flavor-singlet correlation functions in finite volume and for fixed (or imperfectly sampled) topology differs from zero. This has been further investigated numerically in Ref.~\cite{Bali:2014pva}, where the effect on the correlation functions has been shown explicitly for different topological charge sectors. In fact, the leading term in the $1/V$ expansion contributing to at large--$t$ at fixed topological charge $Q$ behaves as
\begin{equation}
 C_\mathrm{2pt}^\mathrm{disc}(t) \sim \frac{a^5}{T}\l(\chi_t - \frac{Q^2}{V} + \frac{c_4}{2V\chi_t}\r)  \,,
 \label{eq:topFS}
\end{equation}
where $\chi_t$ denotes the topological susceptibility and $c_4$ the kurtosis of the topological charge distribution. While we do not find deviations from a zero topological charge in the gauge average on any of the ensembles used in this study, it is still to be expected that an imperfectly sampled topological charge distribution leads to deviations from the infinite volume result. Only on very few ensembles we find a shift at the level of single correlation functions, which is not compatible with zero within errors. This is most notably the case for the light quark correlation function $\tilde{C}_{ll}(t)=\bigl<\tilde{\mathcal{S}}_l^{\tm}(t)\tilde{\mathcal{S}}_l^{\tm}(0)\bigr>$ on D45.32, which yields the dominant contribution to the $\eta'$ principal correlator that is shown in the left panel of Figure~\ref{fig:eigenvalues}. This ensemble exhibits the smallest physical volume of all ensembles in this study. \par

In order to remove any constant shift from our correlation functions we replace the correlators in Eq.~(\ref{eq:corr_matrix}) by a naive time-derivative, i.e. the difference of two adjacent time slices
\begin{equation}
 \tilde{\mathcal{C}}(t) \rightarrow \tilde{\mathcal{C}}(t) - \tilde{\mathcal{C}}(t+\Delta t) \equiv \tilde{\mathcal{C}}'(t) \,, 
 \label{eq:shifted_corr_matrix}
\end{equation}
before solving the generalized eigenvalue problem (GEVP)
\begin{equation}
 \tilde{\mathcal{C}}'(t) v^{n}(t,t_0) = \lambda^{n}(t,t_0) \tilde{\mathcal{C}}'(t_0) v^{n}(t,t_0)
 \label{eq:GEVP}
\end{equation}
where $n=\eta,\eta'$ denotes the two states for the $2\times 2$ problem and using fixed $t_0/a=1$. Taking into account periodic boundary conditions, this changes the functional form of the correlation functions in Eq.~(\ref{eq:corr_matrix}) from an even $\cosh$-like to an odd $\sinh$-like behavior
\begin{equation}
 \tilde{\mathcal{C}}'_{ij}(t) = \sum_n \frac{A^n_i (A^n_j)^*}{2E_n} \cdot 4 \sinh\l(E_n \frac{\Delta t}{2}\r) \exp\l(-E_n \frac{T}{2}\r) \sinh\l(E_n\l(\frac{T-\Delta t}{2}-t\r)\r) \,,
 \label{eq:correlator}
\end{equation}
where $A^n_i$ ($A^n_j$) denotes the physical amplitudes for the $n$th state with respect to the $i$th ($j$th) element of the basis containing $N$ operators. The asymptotic behavior for the principal correlators then takes the same form
\begin{equation}
 \lambda^n(t,t_0) \sim \sinh\l(E_n\l(\frac{T-\Delta t}{2}-t\r)\r) \,,
 \label{eq:eigenvalues}
\end{equation}
which is fitted to the lattice data to extract the energies. Including only correlation functions projected to zero momentum, we have $E_n=M_n$, which yields $\Meta$ and $\Metap$ for the two lowest states. The information on physical amplitudes can be retrieved from eigenvectors in the standard way \cite{Blossier:2009kd}
\begin{equation}
 A^n_i = \frac{\sqrt{2E_n}\cdot\sum_{j=1}^{N} \tilde{C}'_{ij}(t) v^n_j(t,t_0)}{\sqrt{\l(\l(v^n(t,t_0)\r)^T \tilde{C}'(t) v^n(t,t_0)\r) \cdot \lambda^n(t,t_0)}} \,.
 \label{eq:amplitudes}
\end{equation}
Note that amplitudes computed in this way from the correlation function matrix in Eq.~(\ref{eq:corr_matrix}) are renormalized only up to a factor of $Z_P$, which turns out sufficient for our purposes. \par

In addition to removing a constant shift in the $\eta'$ principal correlator, the derivative correlator has much smaller point errors compared to the standard approach, which can be seen comparing the left and the right panel of Figure~\ref{fig:eigenvalues}. The data in the right panel has been generated using a time shift of $\Delta t/a=1$. In fact, the actual choice of $\Delta t$ has little impact on the final results from correlated fits within errors, hence we use $\Delta t/a=1$ throughout our analysis. The only effect of values $\Delta t/a>1$ is less reduction in correlation between adjacent timeslices in $\tilde{C}'(t)$ but in the final fits this is compensated by the behavior of individual point errors. \par

\section{Extraction of mixing parameters and decay widths}
The general definition of meson decay constants employs the axial-vector matrix element in the physical basis
\begin{equation}
 \l< 0 \r| \mathcal{A}^\mu_a \l| P\l(p\r) \r> = i f^P_a p^\mu \,,
 \label{eq:matrix_elements_A}
\end{equation}
where $P=\pi,K,\eta,\eta',...$ denotes the desired meson state (with momentum $p$) and the index $a$ is used to distinguish different flavor structures for the axial-vector current. \par

First we consider the charged meson sector for a light quark doublet imposing exact isospin symmetry. In this case the physical axial-vector current transforms into the vector current in the twisted basis at maximal twist. By virtue of the PCVC relation this leads to a convenient expression for the pion decay constant
\begin{equation}
 f_\pi = 2 \mu_l \frac{\bra{0} \mathcal{P}^a_l \ket{\pi^\pm}}{M_\pi^2} \,, \quad a=1,2 \,,
 \label{eq:f_pi}
\end{equation}
which can be used to compute $f_\pi$ in twisted mass lattice QCD without the need for any renormalization and to high statistical precision due to the pseudoscalar current \cite{Frezzotti:2001du,DellaMorte:2001tu,Jansen:2003ir}. A similar relation holds in the heavy-light meson sector for the kaon
\begin{equation}
 f_K = \l(\mu_l + \mu_s \r)\frac{\bra{0} \tilde{\mathcal{P}}^{+,\tm}_\mathrm{neutral} \ket{K}}{M_K^2} \,,
  \label{eq:f_K}
\end{equation}
where $\tilde{\mathcal{P}}^{+,\tm}_\mathrm{neutral}(x) = \frac{1}{2}\l( Z^{-1} (-\bar{\chi}_d(x)\chi_c(x) + \bar{\chi}_d(x)\chi_s(x)) + \bar{\chi}_d(x) i\g{5} \chi_c(x) + \bar{\chi}_d(x) i\g{5} \chi_s(x)\r) $. While there is again no overall renormalization factor needed, the ratio $Z$ is required due to mixing between scalar and pseudoscalar currents and also implicitly in $\mu_s$ as defined in Eq.~(\ref{eq:heavy_quark_masses}).\par

In the flavor-singlet sector, there exist two popular choices for the basis of two local operators made from degenerate light quark fields and a strange quark field. The first one is the so-called \emph{octet-singlet} basis
\begin{align}
 \mathcal{A}_8^\mu(x)&=\frac{1}{\sqrt{6}} \l(\bar{\psi}_u(x) \G{\mu}\g{5} \psi_u(x) + \bar{\psi}_d(x) \G{\mu}\g{5} \psi_d(x) - 2\bar{\psi}_s(x) \G{\mu}\g{5} \psi_s(x)\r) \,, \\
 \mathcal{A}_0^\mu(x)&=\frac{1}{\sqrt{3}} \l(\bar{\psi}_u(x) \G{\mu}\g{5} \psi_u(x) + \bar{\psi}_d(x) \G{\mu}\g{5} \psi_d(x) +  \bar{\psi}_s(x) \G{\mu}\g{5} \psi_s(x)\r) \,,
\end{align}
which is the preferred basis in the formulation of ($\chi$PT). A second choice is the \emph{quark flavor} basis, defined by
\begin{align}
 \mathcal{A}_l^\mu(x)&=\frac{1}{\sqrt{2}} \l(\bar{\psi}_u(x) \G{\mu}\g{5} \psi_u(x) + \bar{\psi}_d(x) \G{\mu}\g{5} \psi_d(x)\r) \,, \\
 \mathcal{A}_s^\mu(x)&=\bar{\psi}_s(x) \G{\mu}\g{5} \psi_s(x) \,.
\end{align}
In any case, the most general parametrization of the decay constant parameters $f_P^a$ for two local operators made from degenerate light quark fields and a strange quark field, reads:
\begin{equation}
 \l(\begin{array}{ll}
     f^\eta_a  & f^\eta_b \\
     f^{\eta'}_a & f^{\eta'}_b 
    \end{array}\r) = \l(\begin{array}{lr}
     f_a \cos\phi_a & - f_b \sin\phi_b \\
     f_a \sin\phi_a &   f_b \cos\phi_b
    \end{array}\r) \equiv \Xi\l(\phi_a,\phi_b\r) \diag\l(f_a,f_b\r) \,,
 \label{eq:general_mixing}
\end{equation}
where $a=8$, $b=0$ or $a=l$, $b=s$, for octet-singlet and quark flavor basis, respectively. For the octet-singlet basis it is found in $\chi$PT that the leading contribution to the difference of the two mixing angles $\l|\phi_0-\phi_8\r|$ is a purely $\SU{3}_F$--breaking effect, while at the same order in the chiral power counting Okubo-Zweig-Iizuka (OZI)-violating terms contribute only to the flavor-singlet decay constant parameter $f_0 $\cite{schreiber1998proceedings,Feldmann:1998vh}. Therefore $\l|\phi_0-\phi_8\r|$ cannot be expected to be small. On the other hand in the quark flavor basis the corresponding difference $\phi_l-\phi_s$ is proportional only to an OZI-violating term. Since in the $\SU{3}_F$-symmetric theory, the mixing angles fulfill $\phi_s=\phi_l=\arctan\sqrt{2}\equiv\phi_{\SU{3}_F}$, their individual numerical values are not small. This leads to the expectation
\begin{equation}
 \l|\frac{\phi_l-\phi_s}{\phi_l+\phi_s}\r| \ll 1 \,,
\end{equation}
which has been confirmed numerically in a previous lattice study \cite{Michael:2013gka}. Therefore, it is reasonable to define a simplified scheme in the quark flavor basis (the so-called Feldmann-Kroll-Stech scheme \cite{Feldmann:1998vh}), employing only a single mixing angle, i.e. rewriting Eq.~(\ref{eq:general_mixing}) as
\begin{equation}
 \l(\begin{array}{ll}
     f^\eta_l    & f^\eta_s \\
     f^{\eta'}_l & f^{\eta'}_s 
    \end{array}\r) = \Xi\l(\phi,\phi\r) \diag\l(M_\pi^2, 2M_K^2-M_\pi^2\r) \,.
 \label{eq:quark_flavor_mixing}
\end{equation}
For a more detailed discussion of the relation between the two schemes, we refer to the review given in \cite{Feldmann:1999uf}. \par

In principle, the left-hand side of Eq.~(\ref{eq:quark_flavor_mixing}) could be computed directly from the lattice. However, we find that axial-vector interpolating operators do not give a sufficient signal in practice. Therefore, in the physical basis of QCD we resort to pseudoscalar matrix elements
\begin{equation}
 h^P_a= 2 m_a \l<0\r| \mathcal{P}_a \l|P\r> \,,
 \label{eq:matrix_elements_P}
\end{equation}
where again $P=\eta,\eta'$ and $m_a$ for $a=l,s$ denotes the quark mass for the light and strange quark, respectively. The pseudoscalar flavor-singlet operators are the ones given in Eqs.~(\ref{eq:light_op_phys},\ref{eq:heavy_op_phys}), i.e. $\mathcal{P}_l=\mathcal{P}_l^{0,\phys}$ for light quarks and $\mathcal{P}_s=\mathcal{P}_h^{-,\phys}(x)$ for the strange component. Applying $\chi$PT to the same order as for the splitting of the mixing angles, one obtains the desired relation between the mixing parameters from the axial-vector case in Eq.~(\ref{eq:quark_flavor_mixing}) and the matrix elements $h^P_a$ \cite{Feldmann:1999uf}
\begin{equation}
 \l(\begin{array}{ll}
     h^\eta_l    & h^\eta_s \\
     h^{\eta'}_l & h^{\eta'}_s
    \end{array}\r) = \Xi\l(\phi,\phi\r) \diag\l(M_\pi^2 f_l, \l(2M_K^2-M_\pi^2\r) f_s\r) \,.
 \label{eq:pseudoscalar_mixing}
\end{equation}
On the lattice we work in the twisted basis, thus the pseudoscalar currents in the physical basis of QCD need to be replaced by their twisted counterparts. Considering renormalization the matrix elements that are actually computed on the lattice are given by
\begin{align}
 h^{P,\tm,r}_l = \mu_l \l<0\r| \tilde{\mathcal{S}}_l^{\tm} \l|P\r> \,, \label{eq:matrix_elements_P_l_tm} \\ 
 h^{P,\tm,r}_s = \mu_s \l<0\r| \tilde{\mathcal{P}}_s^{\tm} \l|P\r> \,, \label{eq:matrix_elements_P_s_tm}
\end{align}
for $P=\eta,\eta'$. They are obtained from Eq.~(\ref{eq:amplitudes}) solving the GEVP for the correlation function matrix in Eq.~(\ref{eq:corr_matrix}) and multiplying by a factor of the twisted bare light or strange quark mass. Note that the factor $Z_P$, which would otherwise be needed to renormalize the operators $\tilde{\mathcal{S}}_l^{\tm}$, $\tilde{\mathcal{P}}_s^{\tm}$, is canceled by the factor $1/Z_P$ required for the renormalization of the quark masses $\mu_l$, $\mu_s$. We point out that it is an intrinsic advantage of the twisted mass formulation that the computation of the flavor-singlet mixing parameters does not require knowledge of the singlet pseudoscalar renormalization factor $Z_P^0$ at all and that even the nonsinglet $Z_P$ is not required explicitly. The latter is similar to $f_\pi$, which can be computed in the twisted mass formulation without the need for renormalization, or $f_K$, which involves only the ratio of nonsinglet renormalization factors $Z=Z_P/Z_S$. \par
  
Finally, we note that the mixing angle $\phi$ is always invariant under renormalization as it is computed from the (double-) ratio of matrix elements
\begin{equation}
 \tan\phi = -\sqrt{\frac{h^{\eta'}_l h^\eta_s}{h^\eta_l h^{\eta'}_s}} \,,
 \label{eq:mixing_angle}
\end{equation}
unlike the decay constant parameters $f_l$, $f_s$, that depend on $Z$. \par

Using the values for the mixing parameters it is possible to derive estimates and constraints for further physical observables which are driven by the chiral anomaly. First of all, there are relations to the $\eta$ and $\eta'$ decay widths derived from effective field theory, given by~\cite{Feldmann:1998vh}
\begin{align}
 \Getatogg  &= \frac{\alpha_\mathrm{QED}^2 \Meta^3}{288\pi^3}  \cdot \l[\frac{5\cos\phi}{f_l} - \frac{\sqrt{2}\sin\phi}{f_s}\r]^2 \,, \label{eq:Geta2gg} \\
 \Getaptogg &= \frac{\alpha_\mathrm{QED}^2 \Metap^3}{288\pi^3} \cdot \l[\frac{5\sin\phi}{f_l} + \frac{\sqrt{2}\cos\phi}{f_s}\r]^2 \,. \label{eq:Getap2gg} 
\end{align}
Secondly, effective field theory yields relations for the pseudoscalar transition form factors $F_\etagg(q^2)$ and $F_\etapgg(q^2)$ in the limit of large Euclidean momentum transfer $Q^2$ and the mixing parameters in the quark flavor basis \cite{Escribano:2013kba} \footnote{Note that the relative factor of $1/\sqrt{2}$ in our definition compared to Ref.~\cite{Escribano:2013kba} is due to a different normalization of the pion decay constant $f_\pi$, i.e. $92\mev$ vs $130\mev$ (this work).}
\begin{align}
 \limFetagg{Q}  &= \frac{\sqrt{2}}{3} \cdot \l[5 f_l \cos{\phi} - \sqrt{2} f_s \sin{\phi} \r] \equiv \hatFetagg \,, \label{eq:Fetagg} \\
 \limFetapgg{Q} &= \frac{\sqrt{2}}{3} \cdot \l[5 f_l \sin{\phi} + \sqrt{2} f_s \cos{\phi} \r] \equiv \hatFetapgg \,, \label{eq:Fetapgg}
\end{align}
where we have introduced the shorthand notation $\hat{F}_{P\gamma\gamma^*}$, $P=\eta,\eta'$ for later use.

\section{Results}

\begin{table}[t!]
 \centering
  \begin{tabular*}{.8\textwidth}{@{\extracolsep{\fill}}llllll}
    \hline\hline
    ensemble & $aM_\pi$ & $aM_K$ & $a\Meta$ & $a\Metap$ & $\phi \, [\mbox{deg}]$ \\
    \hline\hline
    A30.32   & 0.12384(53) & 0.2511(07) & 0.2807(20) & 0.474(23) & 51.0(2.6) \\
    A40.32   & 0.14128(26) & 0.2569(07) & 0.2847(30) & 0.444(12) & 45.4(1.3) \\
    A40.24   & 0.14520(40) & 0.2590(09) & 0.2859(20) & 0.426(09) & 47.0(1.0) \\
    A60.24   & 0.17316(38) & 0.2663(11) & 0.2930(20) & 0.448(11) & 47.9(0.8) \\
    A80.24   & 0.19922(30) & 0.2779(08) & 0.2945(20) & 0.477(13) & 50.7(1.0) \\
    A100.24  & 0.22161(35) & 0.2878(08) & 0.3034(19) & 0.454(10) & 50.6(0.8) \\
    \hline                             
    A80.24s  & 0.19895(42) & 0.2550(05) & 0.2637(32) & 0.447(19) & 53.2(1.2) \\
    A100.24s & 0.22207(27) & 0.2655(11) & 0.2763(21) & 0.462(11) & 54.6(0.7) \\
    \hline                             
    B25.32   & 0.10708(32) & 0.2130(06) & 0.2373(16) & 0.392(09) & 47.4(1.1) \\
    B35.32   & 0.12530(28) & 0.2181(06) & 0.2405(23) & 0.390(11) & 50.6(1.2) \\
    B55.32   & 0.15567(17) & 0.2288(02) & 0.2481(08) & 0.416(06) & 49.5(0.5) \\
    B75.32   & 0.18082(30) & 0.2378(07) & 0.2500(31) & 0.402(12) & 51.6(1.3) \\
    B85.24   & 0.19299(58) & 0.2459(26) & 0.2493(45) & 0.428(15) & 54.7(1.6) \\
    \hline                             
    D15.48   & 0.06912(30) & 0.1691(12) & 0.1866(35) & 0.298(17) & 39.6(2.7) \\
    D20.48   & 0.07870(26) & 0.1732(03) & 0.1872(50) & 0.346(24) & 38.1(3.2) \\
    D30.48   & 0.09788(29) & 0.1774(04) & 0.1864(46) & 0.319(25) & 39.2(3.7) \\
    D45.32sc & 0.11847(54) & 0.1747(04) & 0.1897(19) & 0.294(11) & 46.1(1.8) \\
    \hline\hline
    \vspace*{0.1cm}
  \end{tabular*}
  \caption{Lattice data for observables that are invariant under renormalization, i.e. mesons masses in lattice units and the mixing angle $\phi$. Note that the kaon mass has been computed only on a smaller subset of configurations, except for A80.24s, B55.32 and D45.32sc. Errors are statistical only.}
  \label{tab:lattice_results_masses_and_angle}
\end{table}

Since our simulations are performed at unphysical values of the quark masses and finite lattice spacing, a chiral extrapolation is required to obtain physical results. To this end we employ a fit ansatz for each observable $O$ inspired by leading order in $\chi$PT
\begin{equation}
\l(r_0^n O[r_0^2\Delta_l, r_0^2\Delta_s, (a/r_0)^2]\r)^m = (r_0^n \chiral{O})^m + \sum_{i=l,s} L_i \cdot r_0^2 \Delta_i + L_\beta \cdot \l(\frac{a}{r_0}\r)^2
\label{eq:fit_ansatz}
\end{equation}
where we defined
\begin{equation}
 \begin{array}{lll}
  \Delta_l =& M_\pi^2 &= 2 B_0 m_l + \mathcal{O}(m^2) \,, \\
  \Delta_s =& 2 M_K^2 - M_\pi^2 &= 2 B_0 m_s + \mathcal{O}(m^2) \,,
 \end{array}
\end{equation}
as leading order proxies for the light and strange quark mass, respectively. In the above fit model $n$ is an integer such that $r_0^n O$ is dimensionless and $m$ denotes the power of the observable required for the chiral expansion, i.e. $m=2$ for masses and $m=1$ for decay constant parameters and the mixing angle $\phi$. The same values of $m$ are used for ratios of the respective quantities, e.g. $m=2$ for fitting $M_\eta/M_K$. The first term on the right-hand side (r.h.s.) of Eq.~(\ref{eq:fit_ansatz}) is included as a free parameter only for observables that do not vanish in the $\SU{3}_F$ chiral limit and take a non-trivial value, such as $M_{\eta'}$, decay constants or ratios thereof. For observables with an analytically known value (e.g. $M_\eta/M_K\rightarrow1$, $\phi\rightarrow\arctan{\sqrt{2}}$) the parameter $\chiral{O}$ is replaced by the respective value. \par

\begin{figure}[t]
  \centering
  \subfigure[]{\includegraphics[width=.48\linewidth]{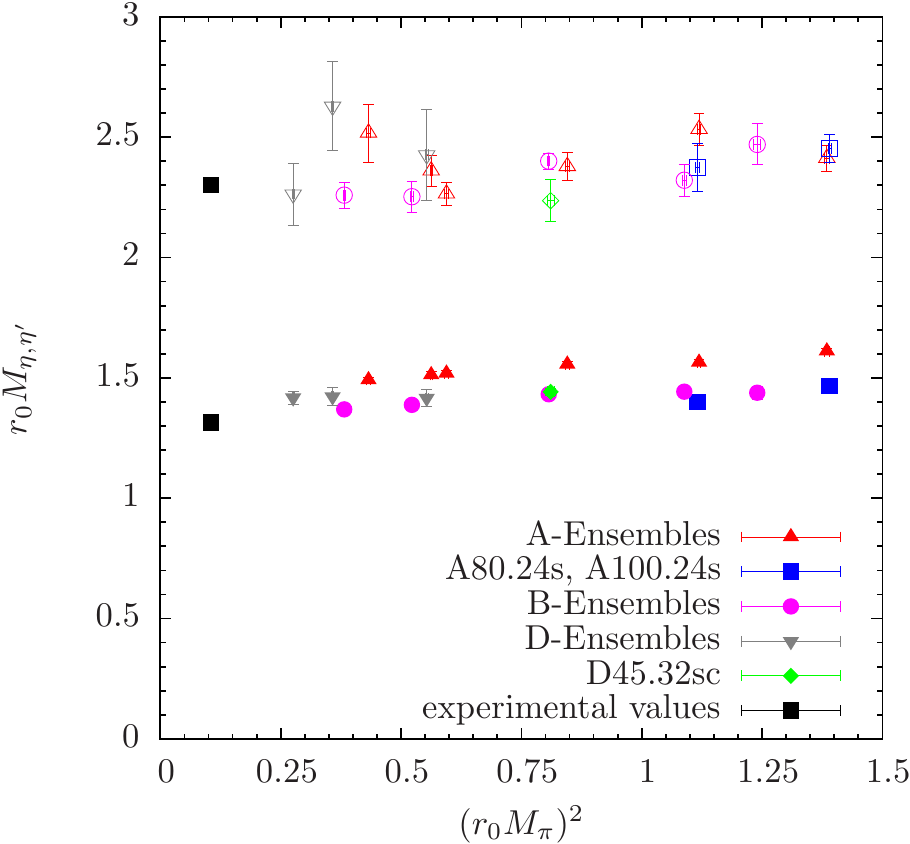}}\quad
  \subfigure[]{\includegraphics[width=.48\linewidth]{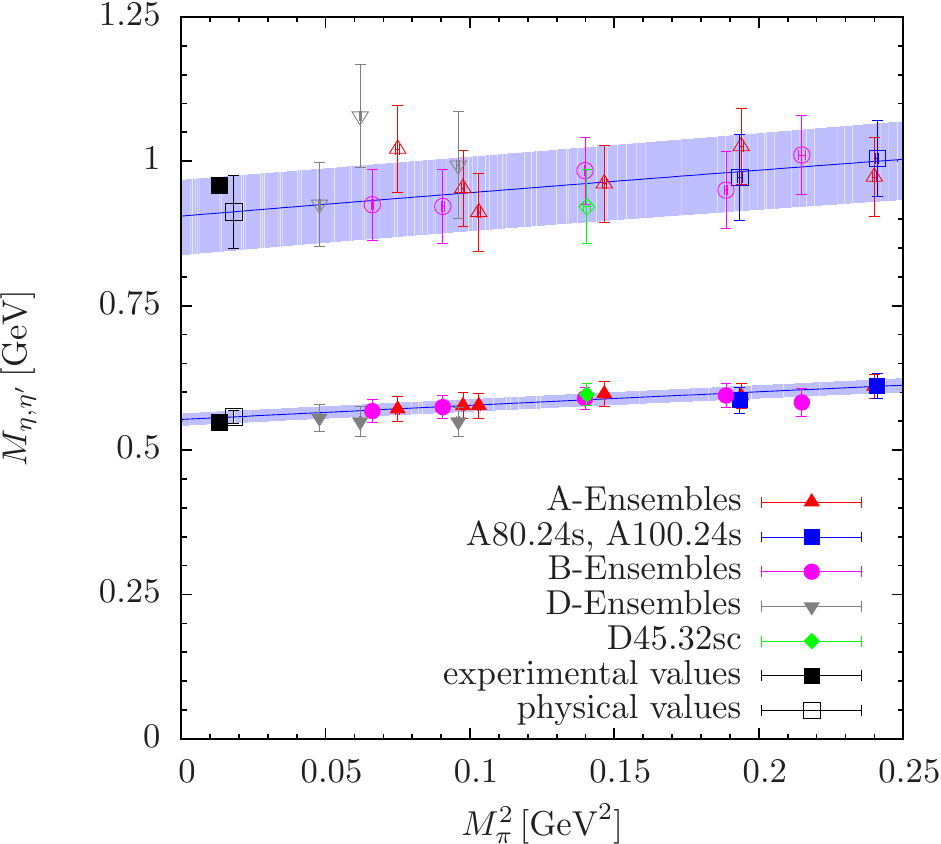}}\quad
  \caption{(a) Measured $\Meta$ (filled symbols) and $\Meta'$ (open symbols) as a function of the pion mass squared in units of $r_0$. Errors on the scale have not been propagated in the data point errors. (b) Combined leading order chiral and continuum extrapolation. Data are corrected to physical value of $m_s$ and for $\mathcal{O}(a^2)$ lattice artifacts using the parameters obtained from the leading order chiral fits. Point errors are increased compared to left panel and highly correlated due to this correction. Conversion to physical units has been done using the value of $r_0$ in Eq.~(\ref{eq:r_0}).}
  \label{fig:masses}
\end{figure}

The constants $L_{l,s,\beta}$ are always free parameters and determined from the fit. They are used to perform chiral and continuum extrapolations, as well as to correct our lattice data for unphysical values of the quark masses and possibly lattice artifacts in plots, e.g. in the right panel of Figure~(\ref{fig:masses}). Note that the resulting point errors are highly correlated. The $\mathcal{O}(a^2)$ term $\sim L_\beta$ has been included to parametrize the leading lattice artifact. \par

The actual observables $O$ considered in our fits are listed in Table~\ref{tab:chiral_fits} together with the resulting fit parameters and values for $\chi^2/\mathrm{dof}$. Note that for the decay constant parameters $f_l$ and $f_s$ we have to resort to fitting ratios $f_l/f_\pi$ and $f_s/f_k$, which cancel most of the lattice artifacts and $m_s$-dependence (in case of $f_s$) that otherwise prevents reasonably fitting the above model.

\begin{table}[t!]
 \centering
 \begin{tabular*}{.8\textwidth}{@{\extracolsep{\fill}}lrrrrll}
  \hline\hline
  $O$  & $(r_0^n \chiral{O})^m$ & $L_l$ & $L_s$ & $L_\beta$ & $\chi^2/\mathrm{dof}$ & $\mathrm{dof}$ \\
  \hline\hline
  $\Meta^2$          & 0                   &  0.280(30) &  0.641(25) &  3.0(2.4)  & 1.31 & 14 \\
  $(\Meta/\MK)^2$    & 1                   & -0.158(17) &  0.094(24) &  1.1(2.3)  & 1.99 & 14 \\
  $\Metap^2$         & 4.1(1.0)            &  0.75(25)  &  0.22(35)  &  7.3(21.3) & 1.35 & 13 \\
  $\phi$             & $\arctan{\sqrt{2}}$ &  0.091(14) & -0.105(12) &  4.5(1.2)  & 1.35 & 14 \\
  $f_l/f_\pi$ \ (M1) & 1.01(06)            & -0.010(20) &  0.015(20) & -6.2(1.4)  & 1.38 & 13 \\
  $f_l/f_\pi$ \ (M2) & 0.96(10)            & -0.040(21) &  0.003(29) & -1.2(1.2)  & 1.39 & 13 \\
  $f_s/f_K$   (M1)   & 1.231(60)           & -0.125(15) &  0.009(26) & -2.6(1.0)  & 1.45 & 13 \\
  $f_s/f_K$   (M2)   & 1.080(54)           & -0.057(18) &  0.025(20) &  1.6(1.0)  & 0.92 & 13 \\
  \hline\hline
  \vspace*{0.1cm}
 \end{tabular*}
 \caption{Final values of the parameters from chiral and continuum fits as defined in Eq.~(\ref{eq:fit_ansatz}) for each observable $O$. For observables with analytically known / trivial value in the chiral limit, the respective parameter $(r_0^n \chiral{O})^m$ has been fixed to this value (values without error) and is not a free parameter in the fit. For the decay constant ratios we include results for both renormalization methods, while all other (invariant) results are for fits to data using $Z$ from M2. The values for the fit parameters of $\phi$ are obtained assuming that $\phi$ and $\chiral{\phi}$ are given in radian measure. Additionally, we include the reduced $\chi^2$ values and degrees of freedom. Errors are statistical only.}
 \label{tab:chiral_fits}
\end{table}

The lattice results for the masses $M_\pi$, $M_K$, $\Meta$ and $\Metap$ are listed together with the mixing angle $\phi$ in Table~\ref{tab:lattice_results_masses_and_angle}. Additional information on the fits of the asymptotic form in Eq.~(\ref{eq:eigenvalues}) to the data for the $\eta$,$\eta'$ principal correlators can be found in the Appendix in Table~\ref{tab:fit_params_final}. For $\Meta$, $\Metap$ and the mixing angle $\phi$ we find that the data are well described by this fit ansatz. In particular the light quark mass dependence for $\Meta$, $\Metap$ is mild over the full range of available pion masses, as can be seen in Figure~\ref{fig:masses}. However, the mass of the $\eta$ depends strongly on the strange quark mass, which is expected from $\chi$PT. Therefore, we consider the ratio with the kaon mass $\Meta/\MK$, which cancels most of the $m_s$ dependence and assign a systematic error from the difference of the two central values. For the remaining observables we assess the uncertainty related to our fitting procedure by performing a cut in the pion mass for the data entering the fits. Including only ensembles with $M_\pi<M_\pi^\mathrm{cut}=390\mev$ reduces the number of data points in the fit from 17 to 11, which is still large enough to obtain reliable fits. We assign a systematic uncertainty to each observable from the difference to the central values from a fit with the aforementioned pion mass cut $M_\pi^\mathrm{cut}=390\mev$, that should reflect the uncertainty related to the chiral extrapolation. Our final results for the $\eta$ and $\eta'$ masses at the physical point read:
\begin{align}
 \Meta^\phys  =& 557(11)_\stat(03)_\sys \mev \,, \\
 \Metap^\phys =& 911(64)_\stat(03)_\sys \mev \,,
 \label{eq:results_masses}
\end{align}
where we have used the experimental values for $\Mpi$ and $\MK$ to set the light and strange quark mass to their physical values and the Sommer parameter $r_0$ in Eq.~(\ref{eq:r_0}) to set the scale. Both results are in good agreement with experiment and compatible with the previous result in Ref.~\cite{Michael:2013gka}. The statistical errors are slightly increased compared to the old results, which is due to the additional degrees of freedom in the now fully consistent, combined chiral and continuum fits. \par

As mentioned before, we have computed the ratio $\Meta/\MK$ to assess the uncertainty of the chiral and continuum extrapolation for the $\eta$ mass. This ratio has indeed been found to cancel most of the strange quark mass dependence in $M_\eta$ \cite{Ottnad:2012fv,Michael:2013gka}. The combined leading order chiral and continuum extrapolation to the physical point using the fit ansatz in Eq.~(\ref{eq:fit_ansatz}) yields
\begin{equation}
 (\Meta/\MK)^\phys = 1.114(31)_\stat \,.
\end{equation}
Plugging in the neutral kaon mass gives $M_\eta=0.554(15)_\stat$, which confirms our result from the direct fit and extrapolation of $M_\eta$. We point out that we use neutral meson masses ($M_{\pi_0}^\experiment$, $M_{K_0}^\experiment$) to set the quark masses and define the physical point, as we do not include electromagnetic effects in our simulations. Using charged meson masses leads to an ambiguity of a few $\mev$, which is still below the statistical uncertainty even for $\Meta$. \par

Regarding further systematics, we have checked for residual excited state effects after replacing the connected correlation functions by the ground state contributions. To this end, we have varied the lower bound of the fit range $t_1^{\eta,\eta'}$. While this yields larger errors for increasing values of $t_1^{\eta,\eta'}$, the results typically agree even within the smaller errors of the fits using $t_1^{\eta,\eta'}=2a$. In particular, there is no significant trend observed from this procedure. Besides, on ensembles with sufficient statistics (e.g. B55.32) we find agreement with results from solving the GEVP without replacing the connected pieces by the ground state contribution. Therefore, it is not possible to resolve any additional excited state contamination within the present statistical precision. \par
 
Similarly, we do not observe significant finite volume effects in the current setup. While the pion and kaon masses in Table~\ref{tab:lattice_results_masses_and_angle} are in principle affected by finite volume effects within their much higher statistical precision, these effects are negligible for the final observables within the statistical error. The remaining observables (i.e. $\Meta$, $\Metap$, $\phi$, $f_{l}$ and $f_s$) are not sensitive to the finite lattice volume within statistical errors. While the ensembles listed in Table~\ref{tab:ensembles} cover different physical volumes as well as different values of $M_\pi L$, there are two dedicated ensembles (A40.24, A40.32) which differ only by their volume, while other physics related parameters are the same for these ensembles. Again, the results on these two ensembles are found to be compatible within errors. \par

\begin{figure}[t]
  \centering
  \subfigure[]{\includegraphics[width=.48\linewidth]{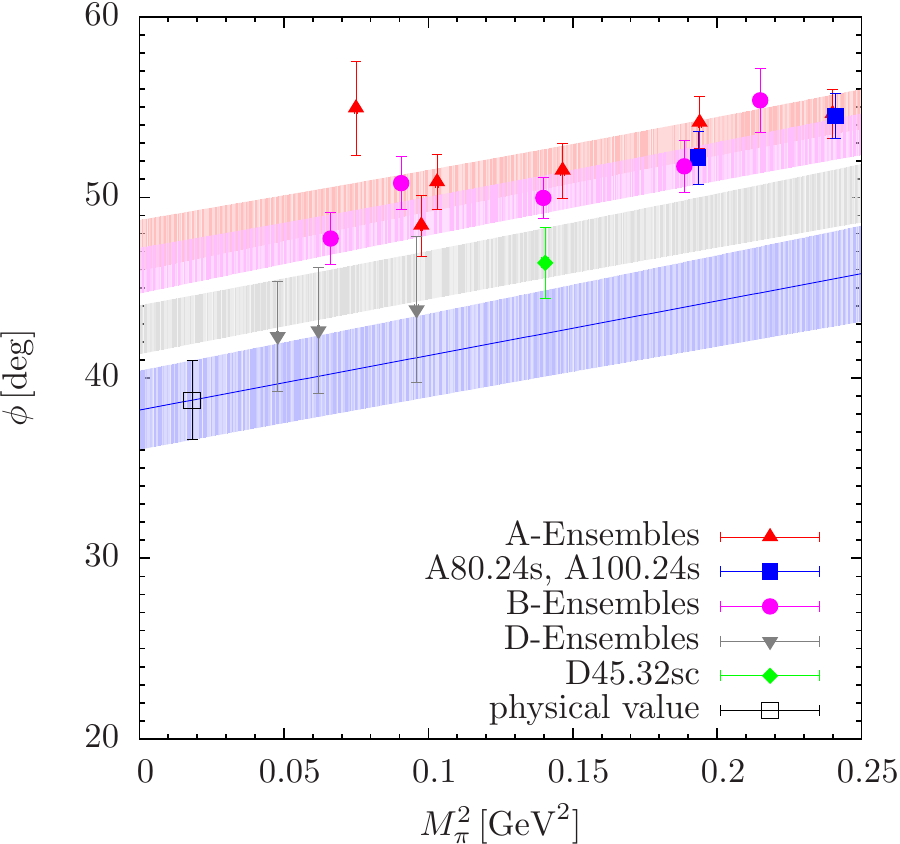}}\quad
  \subfigure[]{\includegraphics[width=.48\linewidth]{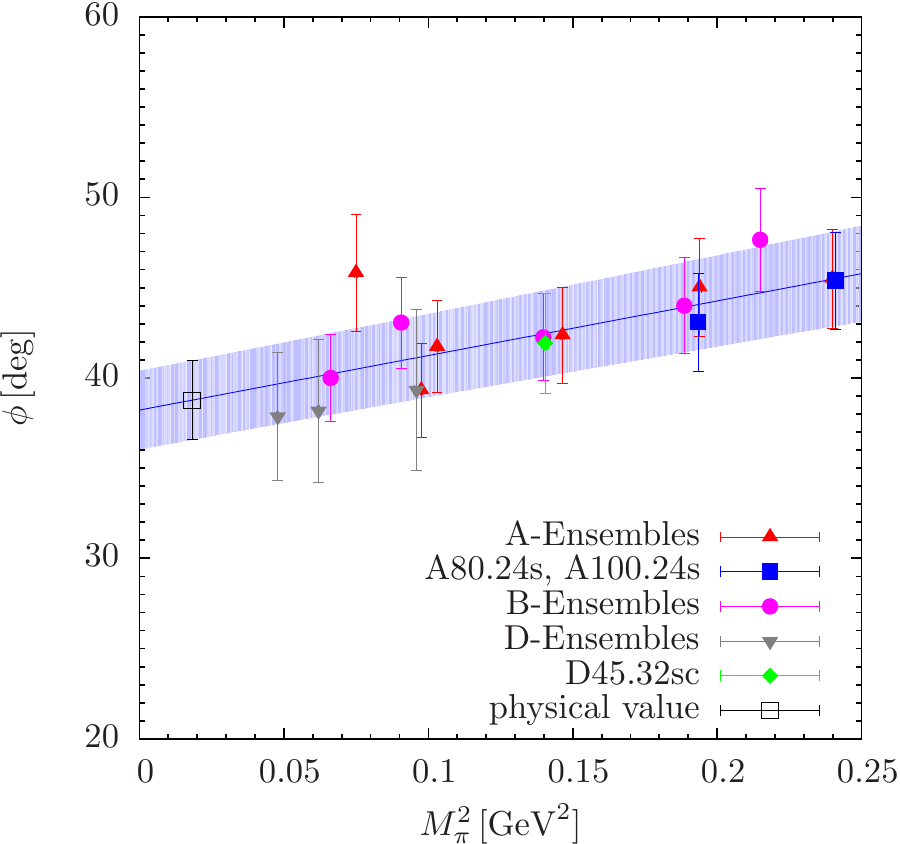}}\quad
  \caption{(a) Mixing angle $\phi$ as a function of $M_\pi^2$ corrected to physical value of the strange quark mass using LO chiral extrapolation. Physical value from full LO fit (see text). The extrapolations resulting from chiral fits at fixed lattice spacing and for physical strange quark mass are shown as red, pink and gray error bands, corresponding to  $\beta$-values of $1.90$, $1.95$ and $2.10$, respectively. Blue error band and solid line correspond to continuum limit extrapolation. (b) Same, but data also corrected for continuum limit. The $\mathrm{SU}(2)$-chiral extrapolation band is shown only for continuum limit and at physical strange quark mass.}
  \label{fig:mixing_angle}
\end{figure}
In Figure~\ref{fig:mixing_angle} we show results from the improved analysis for the mixing angle $\phi$ in the quark flavor basis as defined in Eq.~(\ref{eq:mixing_angle}). The blue band in both panels represents the chiral extrapolation in $M_\pi^2$ in the continuum limit and at physical strange quark mass as obtained from the fit model in Eq.~\ref{eq:fit_ansatz}. While the data in the right panel has been corrected also for the mismatch in the strange quark mass and the continuum limit, the data in the left panel are shown at finite values of the lattice spacing together with an error band from the corresponding chiral extrapolation at fixed lattice spacing. The final result for the mixing angle at physical quark masses and in the continuum reads
\begin{equation}
 \phi^\phys = 38.8(2.2)_\stat(2.4)_\sys^\circ \,,
 \label{eq:result_phi}
\end{equation}
in excellent agreement with results from phenomenology \cite{Feldmann:1999uf,Escribano:2005qq,Escribano:2013kba,Escribano:2015nra,Escribano:2015yup}. We remark that the value quoted above is lower by about three $\sigma$ than what we found in Ref.~\cite{Michael:2013gka}. The reason for this discrepancy is that in Ref.~\cite{Michael:2013gka} we were not sensitive to lattice artifacts. Due to the improved analysis and the additional ensembles, the $a^2$ dependence can be resolved now, which is responsible for a $6^\circ$ decrease in the central value.\par

\begin{figure}[t]
  \centering
  \subfigure[]{\includegraphics[width=.48\linewidth]{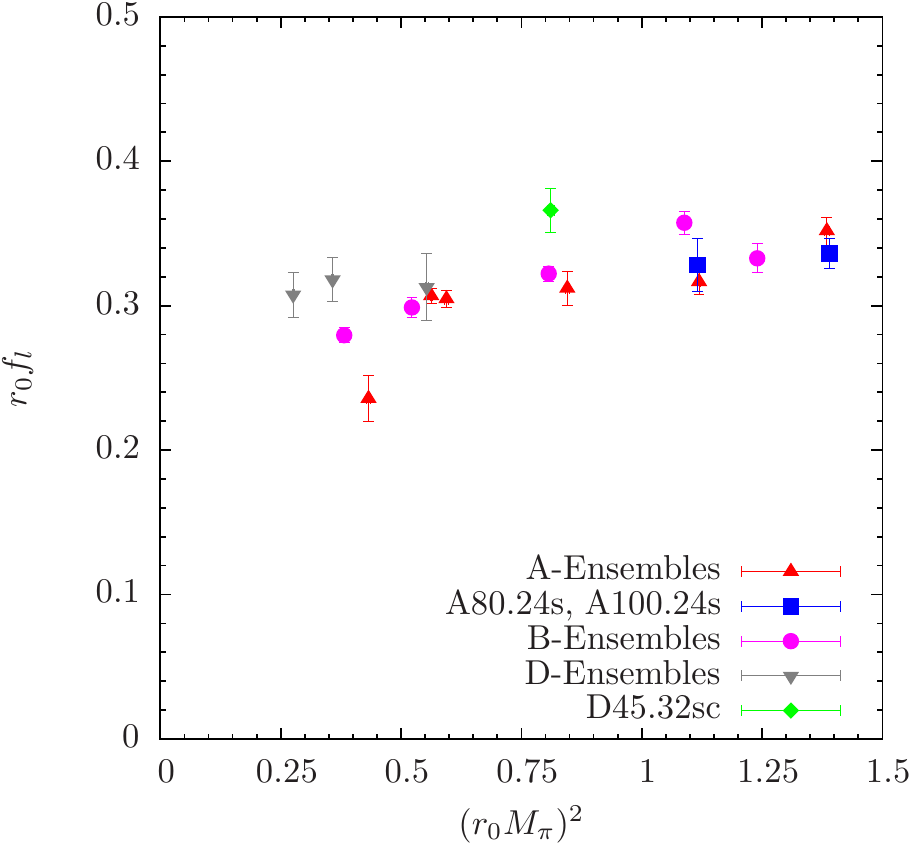}}\quad
  \subfigure[]{\includegraphics[width=.48\linewidth]{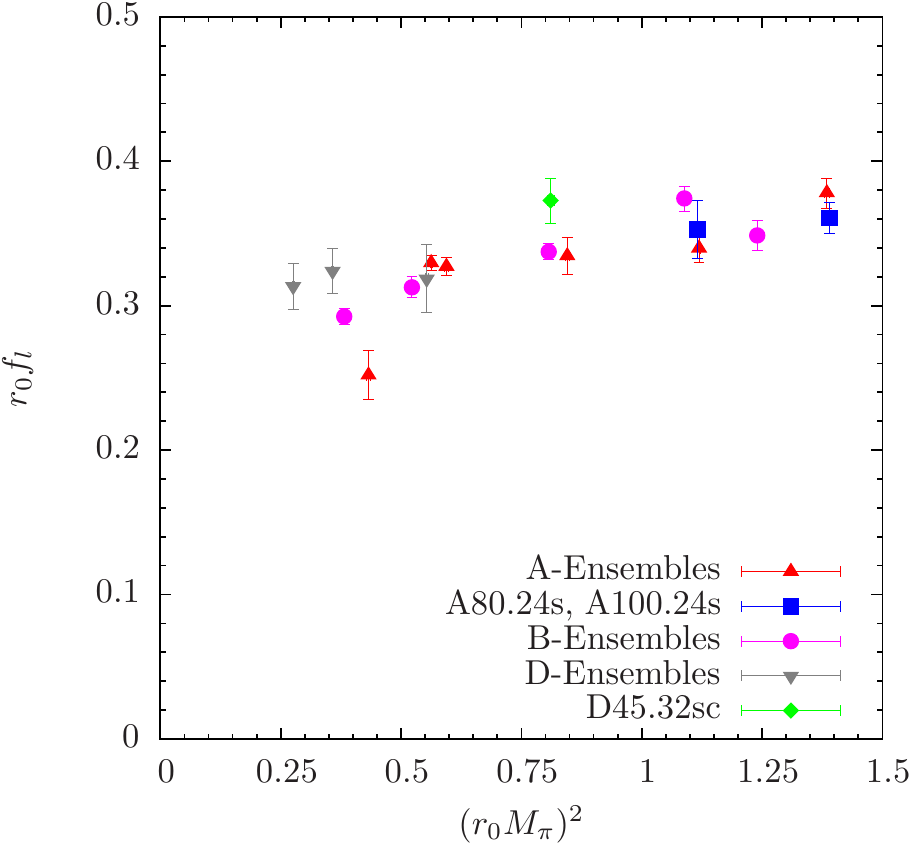}}\quad \\
  \subfigure[]{\includegraphics[width=.48\linewidth]{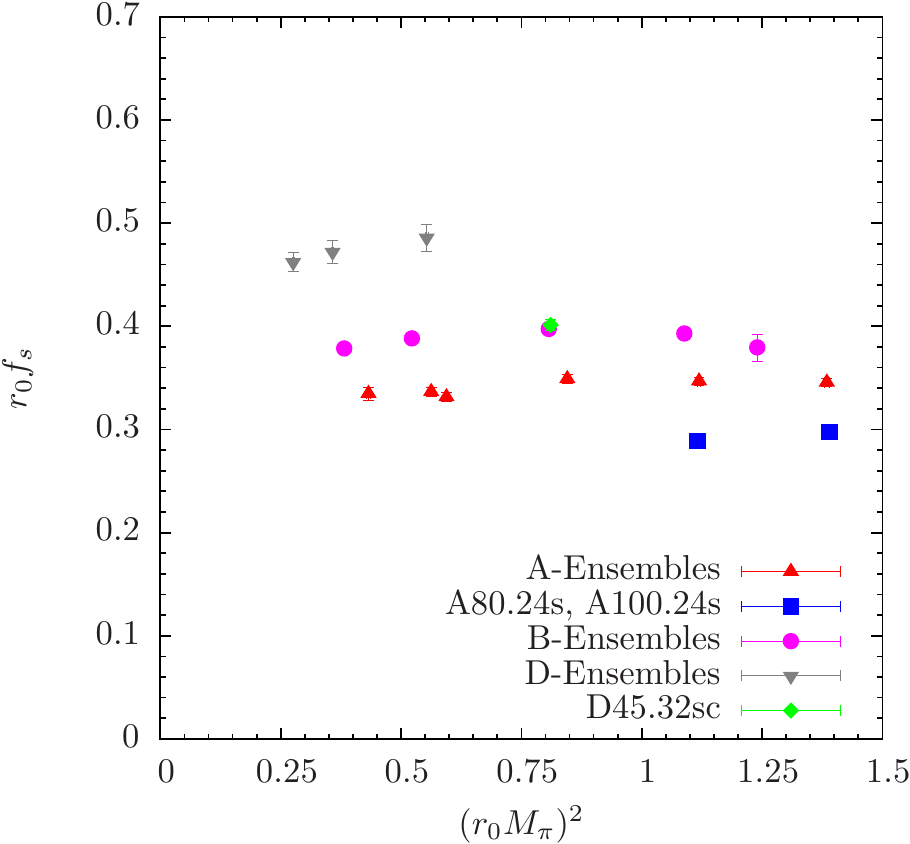}}\quad
  \subfigure[]{\includegraphics[width=.48\linewidth]{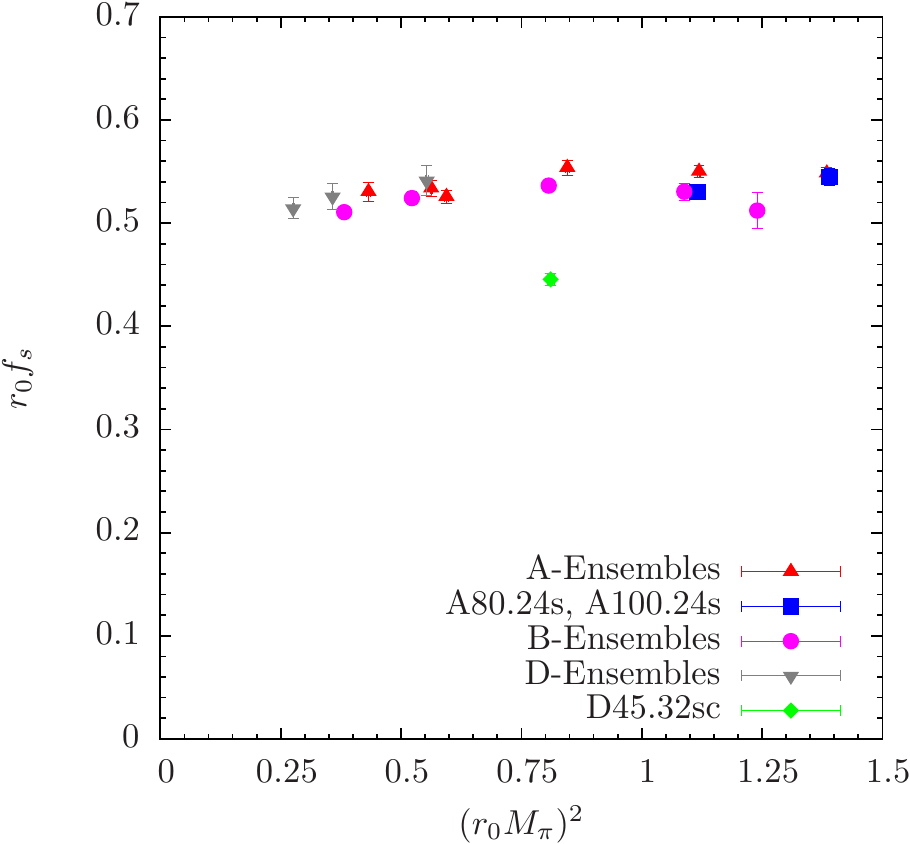}}\quad
  \caption{(a,b) Lattice data for decay constant parameter $f_l$ in units of $r_0$ using $Z$ from renormalization method M1 (left panel) and M2 (right panel). Errors on the scale have not been propagated in the data point errors. (c,d) Same, but for $f_s$.}
  \label{fig:mixing_f}
\end{figure}

In Figure~\ref{fig:mixing_f} we show the lattice data for the decay constant parameters $f_l$ and $f_s$ for both choices of $Z$, i.e. the plots in the left and right columns show data for $Z$ from method M1 and M2, respectively. While $f_l$ is essentially unaffected by the choice of renormalization, the impact on $f_S$ turns out to be very significant. Although formally of $\mathcal{O}(a^2)$, the difference from the choice of renormalization dominates the results and would lead to substantial systematic uncertainties in any attempt of a chiral and continuum extrapolation. Moreover, $f_s$ is rather sensitive to the strange quark mass, as can be inferred from comparing results for ensembles A80.24, A100.24 with their counterparts A80.24s and A100.24s, which have a lighter strange quark mass. This is not surprising, because the matrix element in Eq.~(\ref{eq:matrix_elements_P_s_tm}) that determines $f_s$ is directly proportional to $\mu_s$. However, $\mu_s$ as defined in Eq.~(\ref{eq:heavy_quark_masses}) itself depends explicitly on $Z$. This seems to enhance the effect on $f_s$ of different choices for $Z$. In fact, we cannot exclude that even terms of higher order (e.g. a term $\sim a^2 m_s$) are numerically large for a chiral and continuum extrapolation of $f_s$, hence it is not reasonable to attempt a leading order fit. \par

\begin{figure}[t]
  \centering
  \subfigure[]{\includegraphics[width=.48\linewidth]{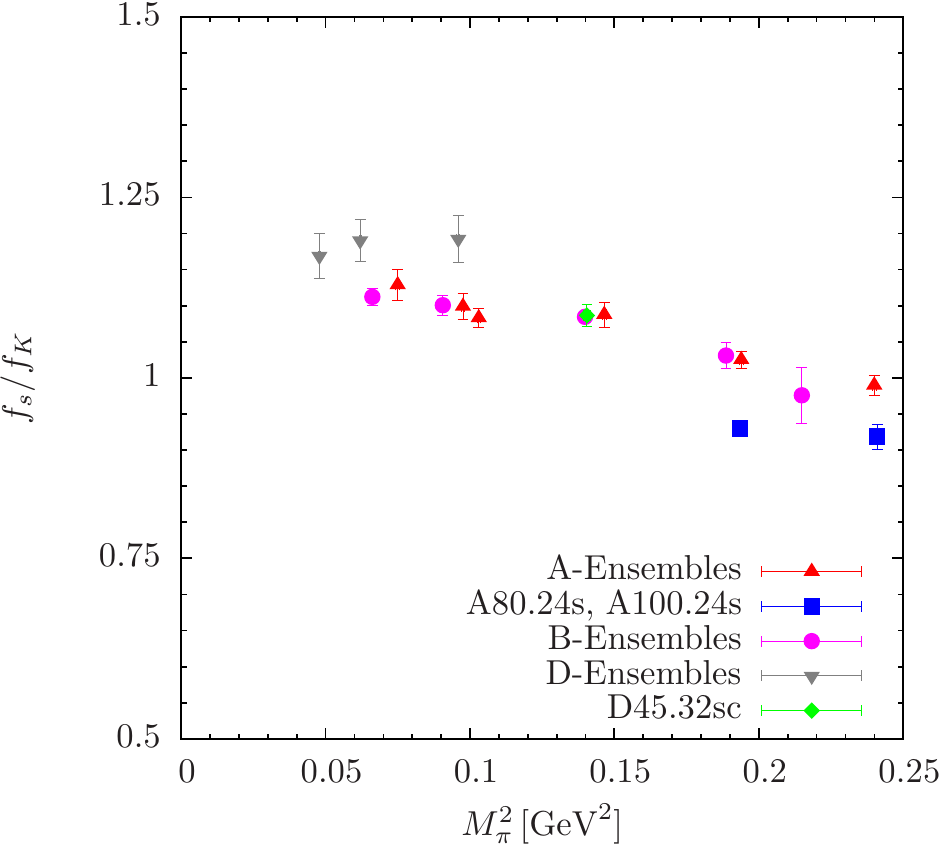}}\quad
  \subfigure[]{\includegraphics[width=.48\linewidth]{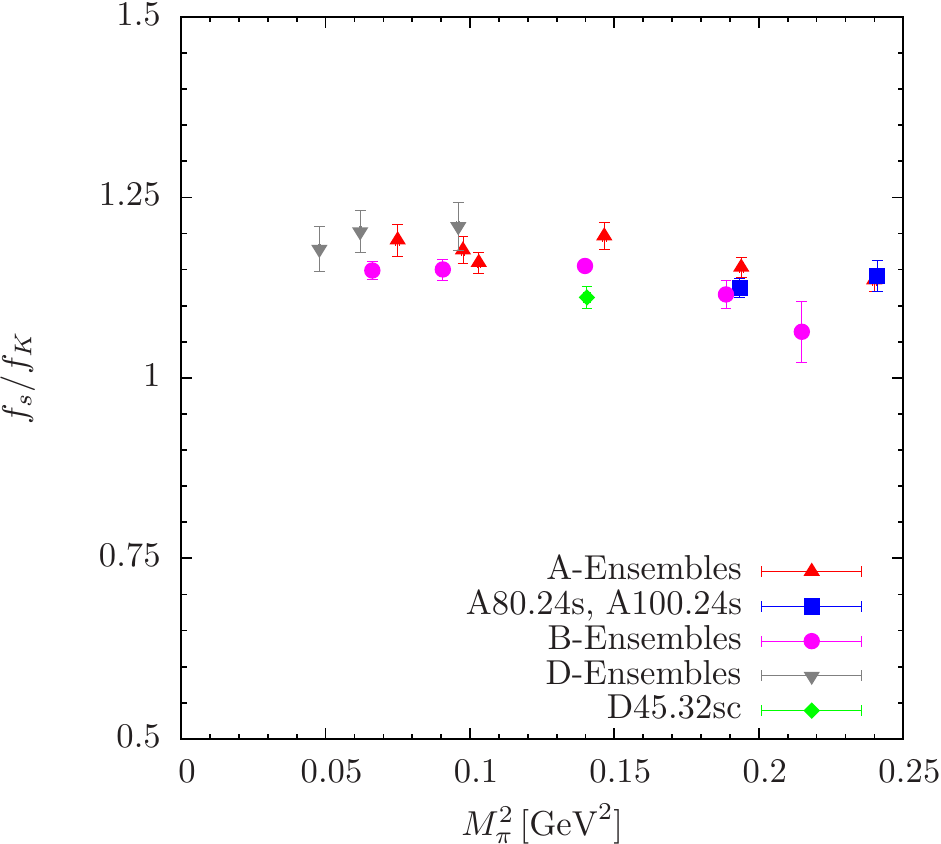}}\quad 
  \caption{(a) Lattice data for $f_s/f_K$ using $Z$ value from method M1. (b) Same but for $Z$ from M2.}
  \label{fig:mixing_f_ratios}
\end{figure}

We find that taking ratios instead of fitting $f_l$, $f_s$ individually allows to circumvent most of these issues and greatly improve the quality of the fits. In particular, we find that forming the ratio of $f_s$ with the kaon decay constant $f_K$ leads to a milder dependence on the choice of $Z$ and it prevents extreme lattice artifacts as observed for $f_s$ using $Z$ from method M1. This is immediately evident from comparing the plots in Figure~\ref{fig:mixing_f_ratios}, which shows results for the ratio $f_s/f_K$ for $Z$ from both methods, with the corresponding ones for $f_s$ in Figure~\ref{fig:mixing_f}. Similarly, taking $f_l/f_\pi$ leads to better fits than considering $f_l$ itself. Still, we find that values for $Z$ computed from method M2 result in generally smaller cut-off effects compared to method M1, as can be seen from the fitted values for $L_\beta$ in Table~\ref{tab:chiral_fits}. Besides, method M2 gives a somewhat better fit for $f_s/f_K$. Therefore, we prefer to choose to use $Z$ from method M2 to compute the final results in our analysis. \par

\begin{figure}[t]
  \centering
  \subfigure[]{\includegraphics[width=.48\linewidth]{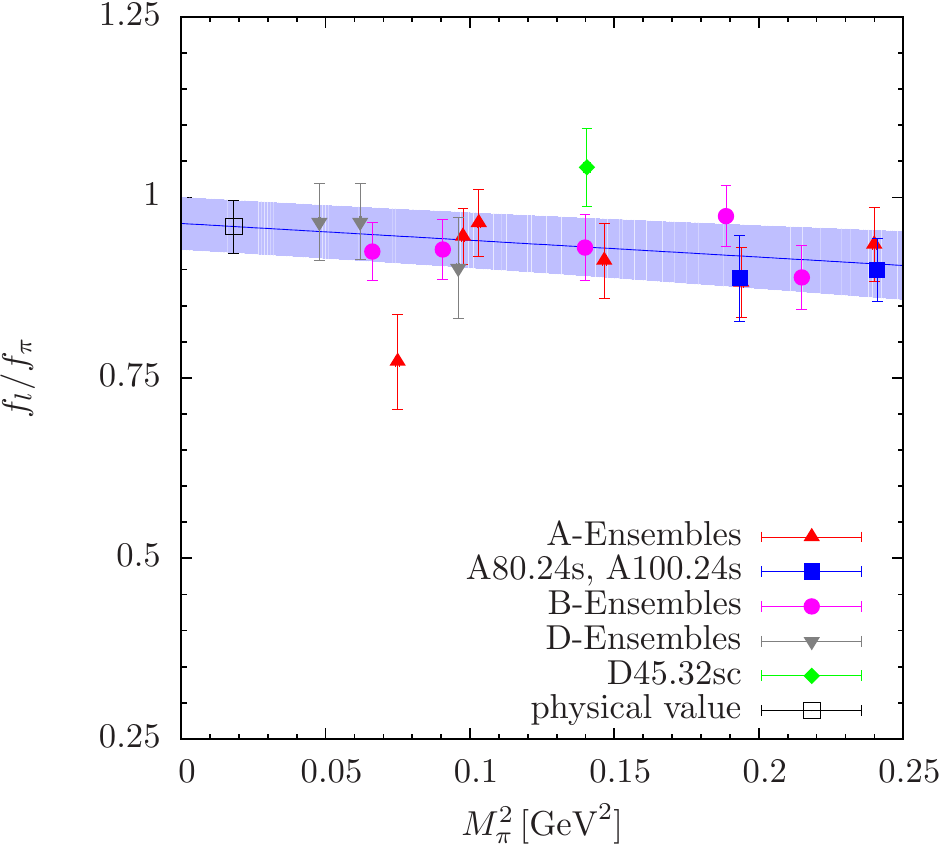}}\quad
  \subfigure[]{\includegraphics[width=.48\linewidth]{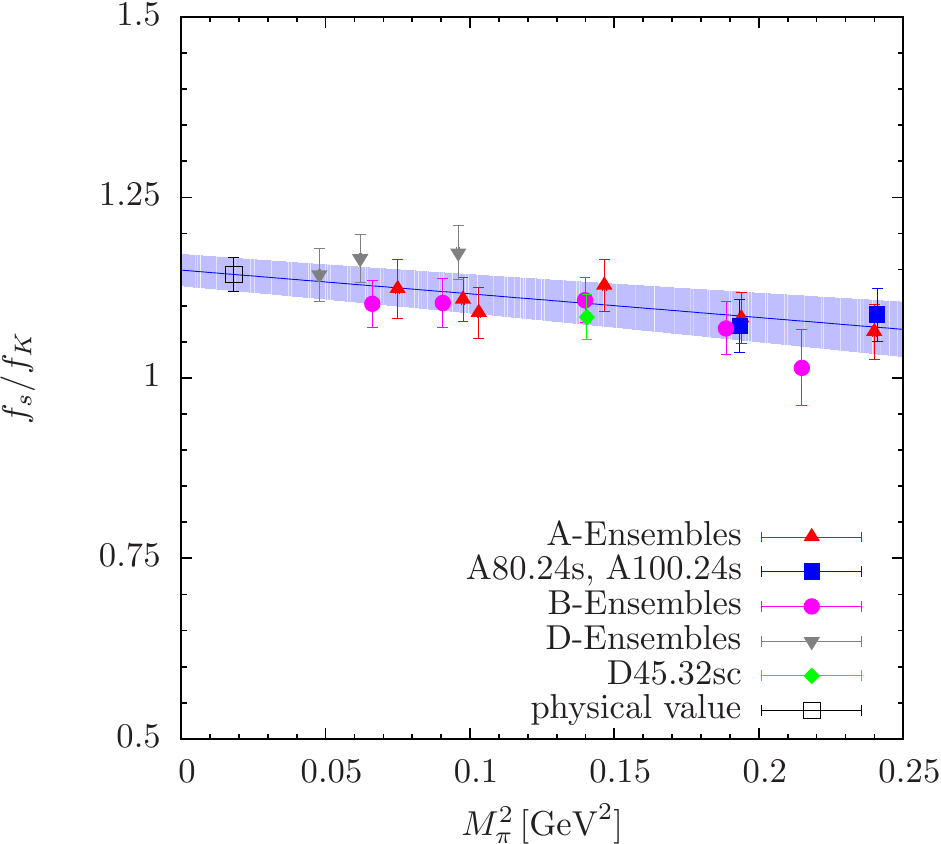}}\quad
  \caption{(a) Lattice data for $f_l/f_\pi$ which has been corrected for the mismatch of the strange quark mass and lattice artifacts from leading order chiral fits. The physical value from the LO fit is shown together with a solid line and gray error band for the chiral extrapolation in $M_\pi^2$ at $m_s=m_{s,\mathrm{phys}}$ and in the continuum limit. (b) Same but for $f_s/f_K$. The data in both panels has been generated using $Z$ from method M2. Errors are statistical only and highly correlated due to the correction for quark masses and the continuum limit.}
  \label{fig:mixing_f_ratios_fitted}
\end{figure}

\begin{table}[t!]
 \centering
  \begin{tabular*}{1\textwidth}{@{\extracolsep{\fill}}lllllll}
    \hline\hline
    ensemble & $af_\pi$   & $af_K$     & $af_l$    & $af_s$    & $f_l/f_\pi$ & $f_s/f_K$ \\
    \hline\hline
    A30.32   & 0.06473(37) & 0.0839(04) & 0.0474(32) & 0.0999(18) & 0.732(49) & 1.190(23) \\
    A40.32   & 0.06851(21) & 0.0854(04) & 0.0621(10) & 0.1005(14) & 0.906(14) & 1.177(19) \\
    A40.24   & 0.06660(31) & 0.0854(05) & 0.0616(12) & 0.0990(12) & 0.925(17) & 1.159(14) \\
    A60.24   & 0.07216(25) & 0.0872(05) & 0.0630(24) & 0.1043(14) & 0.872(34) & 1.196(19) \\
    A80.24   & 0.07593(19) & 0.0898(05) & 0.0640(17) & 0.1036(11) & 0.842(23) & 1.152(14) \\
    A100.24  & 0.07950(18) & 0.0910(05) & 0.0712(19) & 0.1033(11) & 0.895(24) & 1.135(15) \\
    \hline                                         
    A80.24s  & 0.07848(25) & 0.0888(02) & 0.0664(37) & 0.0999(12) & 0.846(47) & 1.125(13) \\
    A100.24s & 0.07915(16) & 0.0899(06) & 0.0679(21) & 0.1026(15) & 0.858(26) & 1.141(22) \\
    \hline                                         
    B25.32   & 0.05695(28) & 0.0771(03) & 0.0507(09) & 0.0885(09) & 0.890(15) & 1.149(12) \\
    B35.32   & 0.06070(23) & 0.0790(04) & 0.0542(13) & 0.0909(11) & 0.893(20) & 1.150(15) \\
    B55.32   & 0.06529(10) & 0.0805(01) & 0.0585(10) & 0.0930(04) & 0.896(15) & 1.155(04) \\
    B75.32   & 0.06906(21) & 0.0824(05) & 0.0649(15) & 0.0919(14) & 0.939(21) & 1.116(20) \\
    B85.24   & 0.07071(35) & 0.0834(11) & 0.0604(18) & 0.0888(30) & 0.854(26) & 1.064(42) \\
    \hline                                         
    D15.48   & 0.04357(23) & 0.0574(07) & 0.0412(21) & 0.0677(13) & 0.947(48) & 1.178(31) \\
    D20.48   & 0.04501(21) & 0.0575(02) & 0.0426(21) & 0.0691(16) & 0.947(46) & 1.203(39) \\
    D30.48   & 0.04747(24) & 0.0588(03) & 0.0419(31) & 0.0712(19) & 0.883(65) & 1.210(33) \\
    D45.32sc & 0.04803(26) & 0.0527(03) & 0.0491(21) & 0.0586(08) & 1.021(41) & 1.111(15) \\
    \hline\hline
    \vspace*{0.1cm}
  \end{tabular*}
  \caption{Results for decay constants and ratios $f_l/f_\pi$, $f_s/f_K$. The $Z$ factors used at each $\beta$ are the ones from method M2, cf. Table~\ref{tab:beta_r0_a_Z}. Note that the kaon decay constant has been computed only on a smaller subset of configurations, except for A80.24s, B55.32 and D45.32sc. Errors are statistical only.}
  \label{tab:lattice_results_decay_constants_M2}
\end{table}

In Table~\ref{tab:lattice_results_decay_constants_M2} we collect the lattice results for $f_l$, $f_s$, $f_l/f_\pi$ and $f_s/f_K$ from method M2 that enter the final fits. Figure~\ref{fig:mixing_f_ratios_fitted} shows the chiral and continuum fit to $f_l/f_\pi$ and $f_s/f_K$ together with the extrapolated lattice data, which appear to be rather well described by the fit ansatz. The final physical results for the decay constant parameters read
\begin{align}
 \l(f_l/f_\pi\r)^\phys =& 0.960(37)_\stat(46)_\sys \quad\mbox{or}\quad f_l^{\phys} = 125(5)_\stat(6)_\sys \mev \,, \\
 \l(f_s/f_K\r)^\phys   =& 1.143(23)_\stat(05)_\sys \quad\mbox{or}\quad f_s^{\phys} = 178(4)_\stat(1)_\sys \mev \,,
 \label{eq:results_decay_constants}
\end{align}
where we have used the experimental values $f_{\pi}^{\experiment}=130.50\mev$, $f_{\mathrm{K}}^{\experiment}=155.72\mev$ to extract $f_l$ and $f_s$, respectively \cite{Olive:2016xmw}. \par

\begin{figure}[t]
  \centering
  \subfigure[]{\includegraphics[width=.48\linewidth]{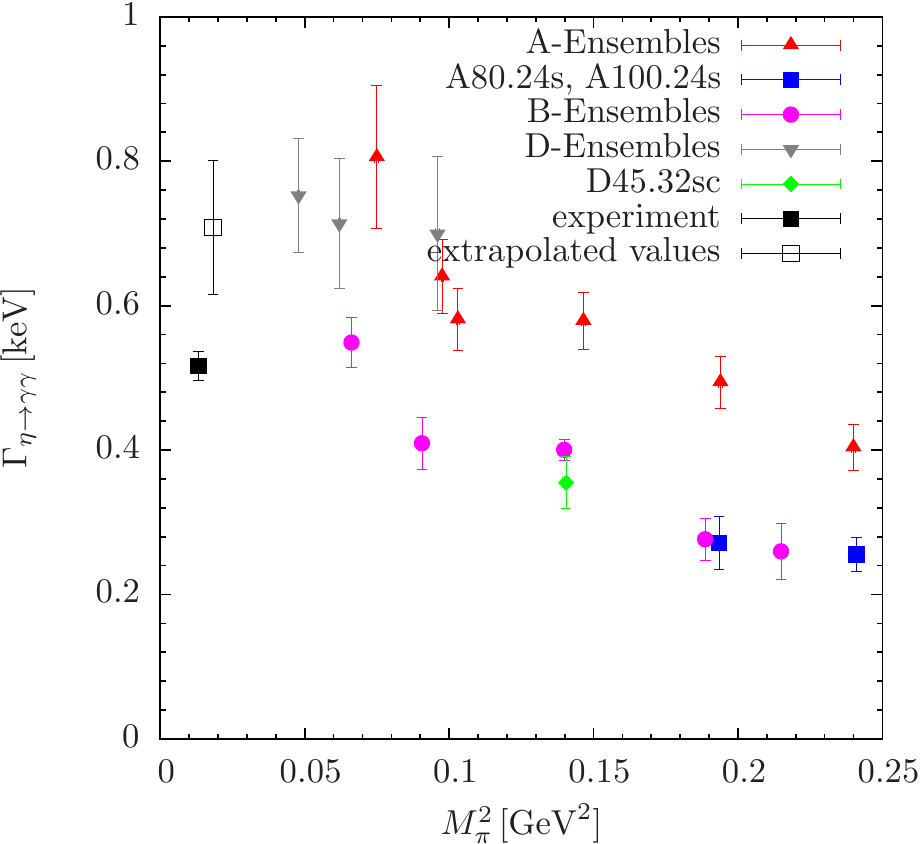}}\quad
  \subfigure[]{\includegraphics[width=.48\linewidth]{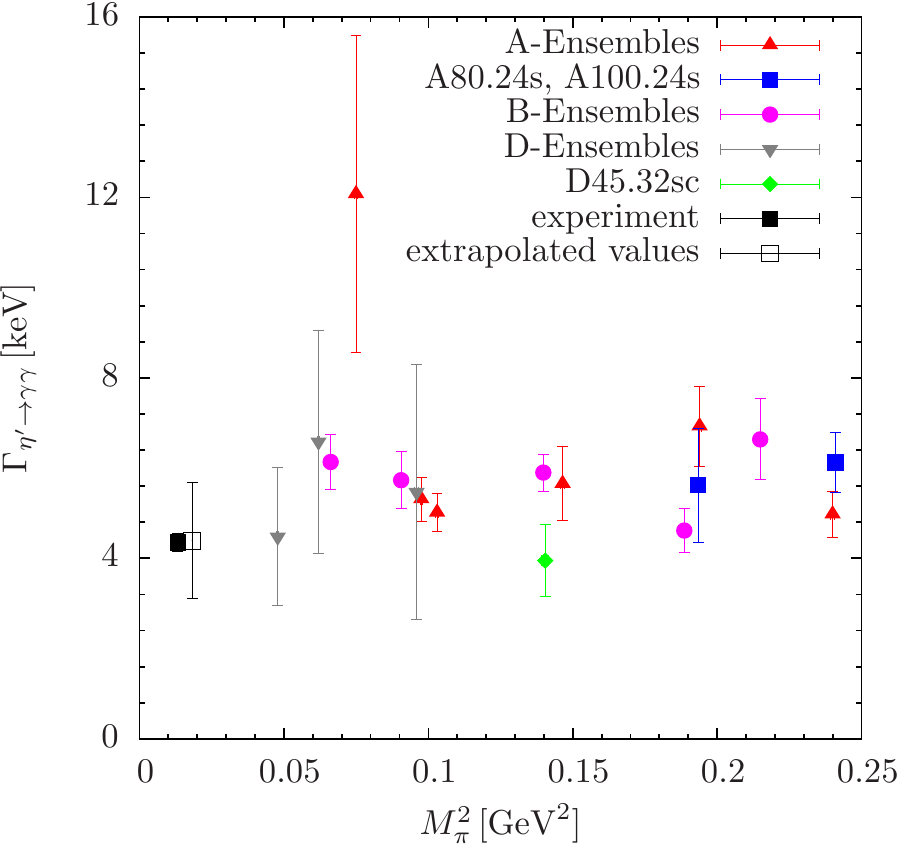}}\quad
  \caption{(a) Results for $\eta\rightarrow\gamma\gamma$ decay widths using $Z$ from method M2. (b) Same, but for $\eta'$. Final, physical values are obtained by plugging in the physical extrapolated values for all the required quantities on the r.h.s. of Eqs.~(\ref{eq:Geta2gg},\ref{eq:Getap2gg}).}
  \label{fig:decay_widths}
\end{figure}

Plugging the physical values for $\Meta$, $\Metap$, $f_l$, $f_s$ and $\phi$ into Eqs.~(\ref{eq:Geta2gg},\ref{eq:Getap2gg}) we can finally compute the $\eta,\eta'\to\gamma\gamma$ decay widths, leading to
\begin{align}
 \Getatogg^\phys  =& 0.71(9)_\stat(7)_\sys \kev \,, \\
 \Getaptogg^\phys =& 4.4(1.3)_\stat(0.6)_\sys \kev \,.
 \label{eq:result_decay_widths}
\end{align}
The large statistical error for $\Getaptogg^\phys$ is dominated by the error on the $\eta'$ mass, which enters to third power in Eq.~(\ref{eq:Getap2gg}). Figure~\ref{fig:decay_widths} shows  $\Getatogg^\phys$ and $\Getaptogg^\phys$ together with the corresponding results at unphysical quark masses and finite lattice spacing, computed on the individual ensembles. Since Eqs.~(\ref{eq:Geta2gg},\ref{eq:Getap2gg}) become rigorous only in the chiral limit, it is expected that there should be corrections at finite quark masses. Indeed, this is clearly observed for $\Getatogg^\phys$, which scales strongly with $m_s$ and exhibits also a residual light quark mass dependence. This might explain why the result differs by more than $2\sigma$ from the PDG value $\Getatogg^\mathrm{exp} = 0.52(2)\kev$ \cite{Olive:2016xmw}. The situation is different for $\Getaptogg^\phys$, which is essentially a constant in $m_l$ and $m_s$, albeit with larger point errors. In fact, the data is compatible with a constant fit over the entire range in $M_\pi$, confirming the applicability of the formula at least for the $\eta'$. The result of such a fit is $\Getaptogg^\phys = 5.5(1.2)_\stat \kev$ with $\chi^2/\mathrm{dof}=20.5/16$. In any case, $\Getaptogg^\phys$ is in agreement with the value $\Getatogg^\mathrm{exp} = 4.4(2)\kev$ within its large error. \par

Similarly, for the transition form factors at large momentum transfer in Eqs.~(\ref{eq:Fetagg},\ref{eq:Fetapgg}) we obtain
\begin{align}
 \hatFetagg^\phys  =& 155(14)_\stat(23)_\sys \mev \,, \\
 \hatFetapgg^\phys =& 277(09)_\stat(01)_\sys \mev \,.
 \label{eq:results_transition_FFs}
\end{align}
The much smaller relative statistical error for $\hatFetapgg$ is caused by anticorrelation, which leads to cancellation of statistical fluctuations in the sum of the two terms on the r.h.s. of Eq.~(\ref{eq:Fetapgg}). In a similar way statistical fluctuations are enhanced in the difference of terms for $\hatFetagg$, while the absolute value of the result is smaller. \par

\begin{figure}[t]
  \centering
  \subfigure[]{\includegraphics[width=.48\linewidth]{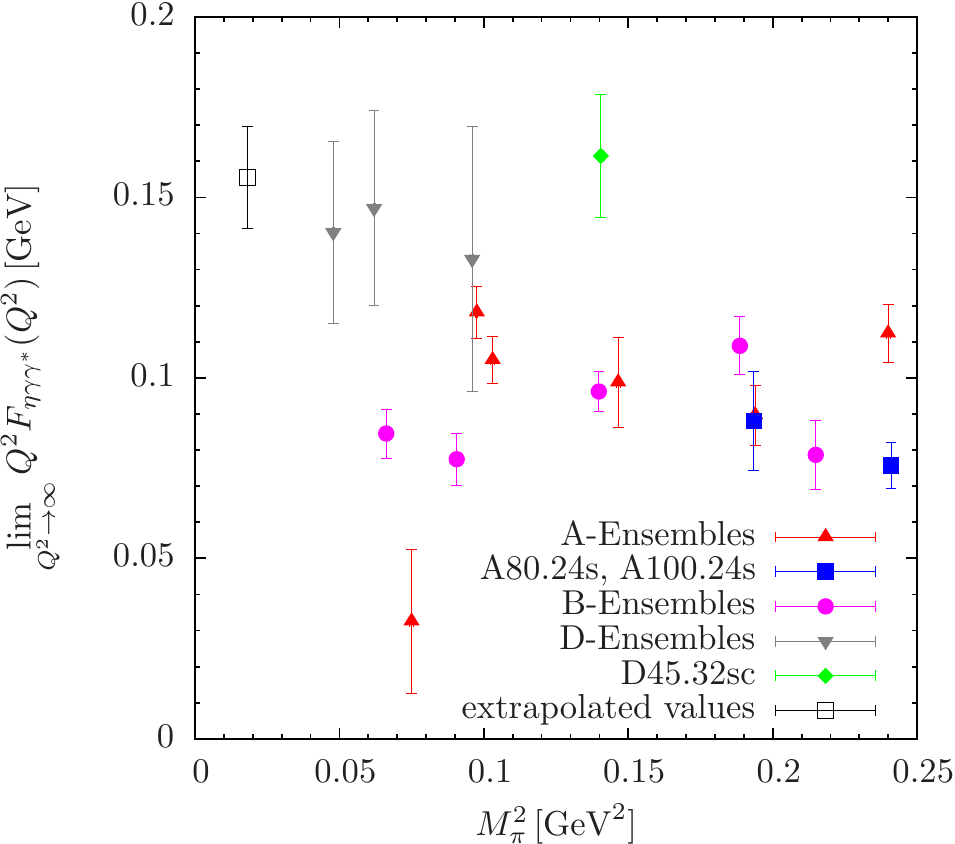}}\quad
  \subfigure[]{\includegraphics[width=.48\linewidth]{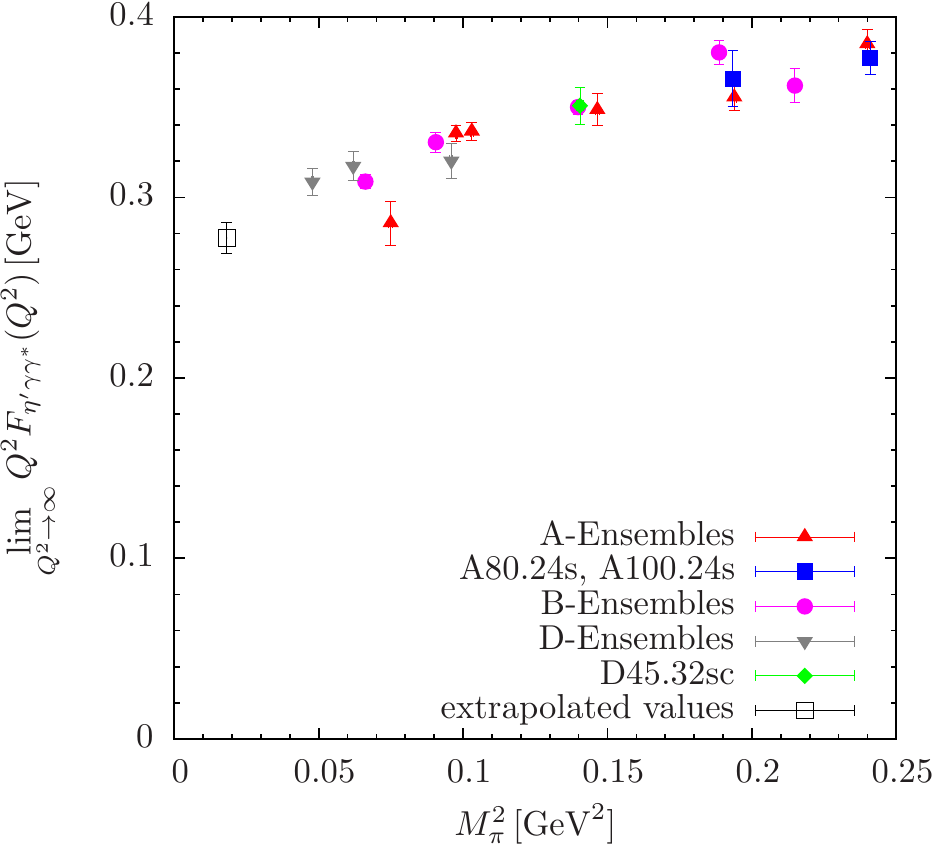}}\quad
  \caption{(a) Results for the $\eta\rightarrow\gamma\gamma$ transition form factor behavior at large $Q^2$ using $Z$ from method M2. (b) Same, but for $\eta'$. Extrapolated (physical) values are obtained by plugging in our physical results for all the required quantities on the r.h.s. of Eqs.~(\ref{eq:Geta2gg},\ref{eq:Getap2gg}).}
  \label{fig:transition_FFs}
\end{figure}

Nevertheless, even for $\hatFetagg^\phys$ the relative statistical precision is better than $10\%$, whereas the systematic uncertainty due to neglecting higher orders in the chiral fits is clearly dominating. The situation seems slightly better for $\hatFetapgg$, but also in this case it is impossible to fully assess the systematics arising due to the use of Eq.~(\ref{eq:Fetapgg}) in the current setup. Therefore, any further improvement beyond the current precision must be subject to a future, dedicated study of $\eta,\eta'\rightarrow\gamma\gamma$ transition form factors, which ultimately should allow to map out the momentum dependence of the transition form factors as well. Still, our results turn out to be in good agreement with a recent phenomenological determination employing a rational approximant analysis in Refs.~\cite{Escribano:2015nra,Escribano:2015yup}. \par

\section{Summary}

In this paper we have for the first time presented results for flavor
singlet, pseudoscalar decay constants using lattice QCD. Thanks to the
gauge ensembles with $N_f=2+1+1$ dynamical quarks provided by the ETMC
collaboration, we could study the continuum and chiral extrapolations
in a controlled way. In particular, dedicated ensembles with varied
bare strange quark mass allow to resolve the strange quark mass
dependence. \par 

For determining the decay constants $f_\ell$ and $f_s$, we had to rely
on $\chi$PT in order to be able to extract them from pseudoscalar matrix
elements. This was necessary, because the axial-vector matrix elements
turned out to be too noisy. While masses and mixing angle(s) are
independent of any renormalization constants, $Z_S$ and $Z_P$ are
needed for $f_\ell$ and $f_s$, which have been determined in two
different ways in Ref.~\cite{Carrasco:2014cwa}. Depending on which way
(M1 or M2) we follow we
observe large lattice artifacts in particular in $f_s$. This we
understand, because in $f_s$ the strange quark mass dependence is
largest and the strange quark mass at separate $\beta$-values is
strongest influenced by the way the renormalization constants are
implemented. Therefore, we decided to rely only on method M2 with
significantly smaller lattice artifacts. \par

Moreover, it turned out that the chiral extrapolation is most
conveniently performed by using ratios $f_s/f_K$ and
$f_\ell/f_\pi$. In those ratios most of the quark mass dependencies
cancel out. Our final estimates for $f_\ell$ and $f_s$ are in very
good agreement to existing phenomenological
determinations~\cite{Feldmann:1999uf}. 
Having $f_\ell$ and $f_s$ at hand we can also -- again relying on $\chi$PT
-- estimate $\eta, \eta'$ decay widths. Within the large
statistical uncertainties we observe reasonable agreement to the PDG
values. \par

Compared to Ref.~\cite{Michael:2013gka}, we have also updated the
results for $M_\eta$, $M_{\eta'}$ and the mixing angle $\phi$ in
the quark flavor basis. This update was necessary because we have
additional ensembles, more statistics and an improved analysis
available. The results are, however, compatible. The somewhat larger
difference in $\phi$ is -- as mentioned already before -- due to the
fact that we can now resolve the lattice spacing dependence. \par

Currently, we are working on improving the signal obtained from the
axial-vector matrix elements. Moreover, we are working on extracting
$\eta$ and $\eta'$ directly at the physical point.
These steps should allow us to cross-check the $\chi$PT formulae used in
this paper and reduce the corresponding systematic uncertainties.

\section*{Acknowledgments}
We would like to thank all members of the European Twisted Mass Collaboration for the most enjoyable collaboration. The computer time for this project was made available to us by the John von Neumann-Institute for Computing (NIC) on the Jugene and Juqueen systems in J{\"u}lich. This project was funded by the DFG as a project in the SFB/TR 16 and the Sino-German CRC110. The authors gratefully acknowledge contributions by Chris Michael at an earlier stage of this project. The gauge configurations used for this work have been generated by the European Twisted Mass Collaboration. The open source software packages tmLQCD~\cite{Jansen:2009xp} and Lemon~\cite{Deuzeman:2011wz} have been used.

\bibliography{refs}
\clearpage

\section{Appendix -- Supplementary tables}
\label{sec:appendix_A}
In the following we collect tables containing additional results and information on the fits performed for this study. Tables~\ref{tab:fit_params_ESRM_00},\ref{tab:fit_params_ESRM_11},\ref{tab:fit_params_ESRM_12},\ref{tab:fit_params_ESRM_21},\ref{tab:fit_params_ESRM_22} contain the value of the lower bound of the fit ranges $[t_1^{ij}/a,..,t_2^{ij}/a]$, $t_2^{ij} = T/2-2a$ to the quark-connected derivative correlators, which are required for replacing them by the respective ground state correlator. Each table corresponds to one matrix elements as indicated in the captions by the indices $i,j=0,...,2$, which are used as a shorthand for labeling the elements of the original $3\times 3$ correlation function matrix made from the operators in Eqs.~(\ref{eq:light_S_tm_eta_operator},\ref{eq:heavy_P_tm_eta_operators}). For technical reasons we perform the replacement of connected contributions at this level, before applying the ratio of renormalization factor $Z$ and rotating to the final operator basis given in Eqs.~(\ref{eq:light_op_renormalized},\ref{eq:strange_op_renormalized}) up to a factor $Z_P$. The charm quark operator is only dropped after this for the actual calculation (i.e. before solving the GEVP), reducing the problem to a $2\times2$ matrix. Note that only five out of nine matrix elements involve a quark-connected contribution to the full correlator. These are the ones entirely made up from either only light or only heavy quarks fields. The contractions for matrix elements, which contain both types of quarks yield only quark-disconnected diagrams. In addition, we include the correlated, reduced $\chi^2$--values ($\chi^2/\mathrm{dof}$) for each fit, together with the corresponding $p$-value and the value of the mass $aM_\mathrm{conn}^{ij}$ and its error, which where used for the identification of the final plateau from scanning different values of $t_1/a$. \par

Table~\ref{tab:fit_params_final} contains the upper bound for the final fit ranges $[t_1^{\eta,\eta'}/a,...,t_2^{\eta,\eta'}/a]$ to the $\eta$ and $\eta'$ principal correlators from solving the GEVP in Eq.~(\ref{eq:GEVP}). Besides, the resulting correlated $\chi^2/\mathrm{dof}$ and $p$--values are given. \par

Finally, in Table~\ref{tab:lattice_results_P2gg_M2} we have included the numerical data for the decay widths and transition form factors for each ensemble, which have been used for Figure~\ref{fig:decay_widths} and Figure~\ref{fig:transition_FFs}, respectively. \par

\begin{table}[hbt]
 \centering
  \begin{tabular*}{.8\textwidth}{@{\extracolsep{\fill}}lrrrr}
    \hline\hline
    ensemble & $t_1^{00}/a$ & $(\chi^2/\mathrm{dof})^{00}$ & $p$-value & $aM_\mathrm{conn}^{00}$ \\
    \hline\hline
    A30.32   & 18 & 1.31 & 0.214 & 0.2095(66) \\
    A40.32   & 10 & 1.02 & 0.435 & 0.2301(21) \\
    A40.24   &  9 & 0.94 & 0.501 & 0.2402(29) \\
    A60.24   & 14 & 0.88 & 0.518 & 0.2488(58) \\
    A80.24   & 15 & 1.34 & 0.235 & 0.2576(36) \\
    A100.24  & 16 & 0.62 & 0.686 & 0.2851(34) \\
    \hline                 
    A80.24s  & 18 & 0.85 & 0.468 & 0.2510(78) \\
    A100.24s & 16 & 0.88 & 0.495 & 0.2774(35) \\
    \hline                 
    B25.32   & 11 & 0.96 & 0.507 & 0.1813(21) \\
    B35.32   & 14 & 1.50 & 0.095 & 0.1907(21) \\
    B55.32   & 19 & 0.93 & 0.504 & 0.2117(15) \\
    B75.32   & 18 & 0.93 & 0.512 & 0.2350(21) \\
    B85.24   & 13 & 1.21 & 0.289 & 0.2314(35) \\
    \hline                 
    D15.48   & 28 & 1.64 & 0.045 & 0.1108(34) \\
    D20.48   & 27 & 2.17 & 0.003 & 0.1170(32) \\
    D30.48   & 32 & 1.08 & 0.370 & 0.1275(40) \\
    D45.32sc & 21 & 0.89 & 0.528 & 0.1534(38) \\
    \hline\hline
    \vspace*{0.1cm}
  \end{tabular*}
  \caption{Fit parameters and resulting masses with statistical errors for ground-state fits to the quark-connected correlation function on matrix element $i=j=0$, i.e. $\bigl<\mathcal{S}_l^{3,\tm}(t) \mathcal{S}_l^{3,\tm}(0)\bigr>_\mathrm{conn}$.}
  \label{tab:fit_params_ESRM_00}
\end{table}

\begin{table}[t!]
 \centering
  \begin{tabular*}{.8\textwidth}{@{\extracolsep{\fill}}lllll}
    \hline\hline
    ensemble & $t_1^{11}/a$ & $\chi^2/\mathrm{dof}$ & $p$-value & $aM_\mathrm{conn}^{11}$ \\
    \hline\hline
    A30.32   & 20 & 0.96 & 0.475 & 0.30683(07) \\
    A40.32   & 23 & 0.87 & 0.518 & 0.30808(10) \\
    A40.24   & 15 & 0.80 & 0.567 & 0.30756(10) \\
    A60.24   & 11 & 0.80 & 0.630 & 0.31136(08) \\
    A80.24   & 10 & 0.69 & 0.763 & 0.31580(06) \\
    A100.24  & 16 & 0.89 & 0.487 & 0.31664(09) \\
    \hline
    A80.24s  & 14 & 2.15 & 0.035 & 0.27159(10) \\
    A100.24s & 12 & 0.91 & 0.518 & 0.27604(07) \\
    \hline
    B25.32   & 16 & 1.17 & 0.292 & 0.25747(05) \\
    B35.32   & 15 & 0.96 & 0.491 & 0.26256(05) \\
    B55.32   & 19 & 0.92 & 0.512 & 0.26426(03) \\
    B75.32   & 14 & 0.98 & 0.473 & 0.26317(04) \\
    B85.24   & 12 & 0.99 & 0.442 & 0.26409(12) \\
    \hline
    D15.48   & 34 & 1.36 & 0.184 & 0.22205(05) \\
    D20.48   & 32 & 0.95 & 0.494 & 0.22308(05) \\
    D30.48   & 36 & 1.13 & 0.339 & 0.22467(07) \\
    D45.32sc & 20 & 0.97 & 0.459 & 0.21052(09) \\
    \hline\hline
    \vspace*{0.1cm}
  \end{tabular*}
  \caption{Fit parameters and resulting masses with statistical errors for ground-state fits to the quark-connected correlation function on matrix element $i=j=1$, i.e. $\bigl<\mathcal{P}_h^{+,\tm}(t) \mathcal{P}_h^{+,\tm}(0)\bigr>_\mathrm{conn}$.}
  \label{tab:fit_params_ESRM_11}
\end{table}

\begin{table}[t!]
 \centering
  \begin{tabular*}{.8\textwidth}{@{\extracolsep{\fill}}lllll}
    \hline\hline
    ensemble & $t_1^{12}/a$ & $\chi^2/\mathrm{dof}$ & $p$-value & $aM_\mathrm{conn}^{12}$ \\
    \hline\hline
    A30.32   & 21 & 0.92 & 0.497 & 0.3068(08)  \\
    A40.32   & 23 & 1.00 & 0.425 & 0.3077(09)  \\
    A40.24   & 14 & 0.89 & 0.517 & 0.3079(09)  \\
    A60.24   & 11 & 1.42 & 0.163 & 0.3115(08)  \\
    A80.24   & 10 & 0.97 & 0.475 & 0.3161(06)  \\
    A100.24  & 16 & 0.89 & 0.485 & 0.3166(09)  \\
    \hline
    A80.24s  & 16 & 2.34 & 0.039 & 0.2728(13)  \\
    A100.24s & 17 & 0.88 & 0.474 & 0.2766(06)  \\
    \hline
    B25.32   & 21 & 1.65 & 0.106 & 0.2579(08)  \\
    B35.32   & 15 & 0.94 & 0.512 & 0.2623(05)  \\
    B55.32   & 20 & 0.96 & 0.471 & 0.2642(03)  \\
    B75.32   & 24 & 0.87 & 0.502 & 0.2626(14)  \\
    B85.24   & 13 & 0.91 & 0.507 & 0.2643(13)  \\
    \hline
    D15.48   & 33 & 1.47 & 0.127 & 0.2219(04)  \\
    D20.48   & 36 & 0.71 & 0.702 & 0.2247(06)  \\
    D30.48   & 36 & 0.88 & 0.540 & 0.2245(07)  \\
    D45.32sc & 23 & 0.91 & 0.489 & 0.2092(15)  \\
    \hline\hline
    \vspace*{0.1cm}
  \end{tabular*}
  \caption{Fit parameters and resulting masses with statistical errors for ground-state fits to the quark-connected correlation function on matrix element $i=1$, $j=2$, i.e. $\bigl<\mathcal{P}_h^{+,\tm}(t) \mathcal{P}_h^{-,\tm}(0)\bigr>_\mathrm{conn}$.}
  \label{tab:fit_params_ESRM_12}
\end{table}

\begin{table}[t!]
 \centering
  \begin{tabular*}{.8\textwidth}{@{\extracolsep{\fill}}lllll}
    \hline\hline
    ensemble & $t_1^{21}/a$ & $\chi^2/\mathrm{dof}$ & $p$-value & $aM_\mathrm{conn}^{21}$ \\
    \hline\hline
    A30.32   & 24 & 0.71 & 0.618 & 0.3074(10) \\
    A40.32   & 24 & 1.12 & 0.349 & 0.3088(10) \\
    A40.24   & 16 & 0.81 & 0.542 & 0.3079(10) \\
    A60.24   & 11 & 0.83 & 0.601 & 0.3116(07) \\
    A80.24   & 11 & 1.21 & 0.279 & 0.3154(06) \\
    A100.24  & 17 & 0.86 & 0.484 & 0.3162(10) \\
    \hline
    A80.24s  & 14 & 2.22 & 0.299 & 0.2719(09) \\
    A100.24s & 12 & 0.96 & 0.473 & 0.2759(06) \\
    \hline
    B25.32   & 17 & 1.56 & 0.971 & 0.2580(06) \\
    B35.32   & 16 & 0.96 & 0.486 & 0.2626(04) \\
    B55.32   & 19 & 0.81 & 0.620 & 0.2643(03) \\
    B75.32   & 22 & 0.92 & 0.486 & 0.2622(08) \\
    B85.24   & 12 & 1.22 & 0.279 & 0.2648(11) \\
    \hline
    D15.48   & 32 & 1.30 & 0.204 & 0.2222(03) \\
    D20.48   & 34 & 0.96 & 0.479 & 0.2238(05) \\
    D30.48   & 36 & 1.14 & 0.328 & 0.2248(06) \\
    D45.32sc & 23 & 0.86 & 0.527 & 0.2098(13) \\
    \hline\hline
    \vspace*{0.1cm}
  \end{tabular*}
  \caption{Fit parameters and resulting masses with statistical errors for ground-state fits to the quark-connected correlation function on matrix element $i=2$, $j=1$, i.e. $\bigl<\mathcal{P}_h^{-,\tm}(t) \mathcal{P}_h^{+,\tm}(0)\bigr>_\mathrm{conn}$.}
  \label{tab:fit_params_ESRM_21}
\end{table}

\begin{table}[t!]
 \centering
  \begin{tabular*}{.8\textwidth}{@{\extracolsep{\fill}}lllll}
    \hline\hline
    ensemble & $t_1^{22}/a$ & $\chi^2/\mathrm{dof}$ & $p$-value & $aM_\mathrm{conn}^{22}$ \\
    \hline\hline
    A30.32   & 20 & 0.95 & 0.477 & 0.30680(08)  \\
    A40.32   & 24 & 0.76 & 0.578 & 0.30890(10)  \\
    A40.24   & 15 & 0.91 & 0.490 & 0.30834(09)  \\
    A60.24   & 18 & 0.30 & 0.825 & 0.31221(19)  \\
    A80.24   & 18 & 0.37 & 0.774 & 0.31463(13)  \\
    A100.24  & 12 & 0.92 & 0.506 & 0.31742(05)  \\
    \hline
    A80.24s  & 16 & 2.23 & 0.048 & 0.27287(11)  \\
    A100.24s & 14 & 1.01 & 0.425 & 0.27622(08)  \\
    \hline
    B25.32   & 21 & 1.93 & 0.506 & 0.25792(07)  \\
    B35.32   & 23 & 0.88 & 0.511 & 0.26089(09)  \\
    B55.32   & 21 & 0.91 & 0.503 & 0.26414(03)  \\
    B75.32   & 23 & 0.83 & 0.548 & 0.26205(10)  \\
    B85.24   & 12 & 1.06 & 0.388 & 0.26489(11)  \\
    \hline
    D15.48   & 33 & 1.41 & 0.154 & 0.22219(04)  \\
    D20.48   & 36 & 0.72 & 0.692 & 0.22486(06)  \\
    D30.48   & 36 & 0.87 & 0.555 & 0.22460(06)  \\
    D45.32sc & 24 & 0.66 & 0.657 & 0.20932(16)  \\
    \hline\hline
    \vspace*{0.1cm}
  \end{tabular*}
  \caption{Fit parameters and resulting masses with statistical errors for ground-state fits to the quark-connected correlation function on matrix element $i=j=2$, i.e. $\bigl<\mathcal{P}_h^{-,\tm}(t) \mathcal{P}_h^{-,\tm}(0)\bigr>_\mathrm{conn}$.}
  \label{tab:fit_params_ESRM_22}
\end{table}

\begin{table}[t!]
 \centering
  \begin{tabular*}{.8\textwidth}{@{\extracolsep{\fill}}lllllll}
    \hline\hline
    ensemble & $t_2^{\eta}/a$ & $t_2^{\eta'}/a$ & $(\chi^2/\mathrm{dof})^{\eta}$ & $(\chi^2/\mathrm{dof})^{\eta'}$ & $p^{\eta}$ & $p^{\eta'}$ \\
    \hline\hline
    A30.32   & 14 &  8 & 0.92 & 0.96 & 0.518 & 0.443 \\
    A40.32   &  8 &  8 & 1.71 & 1.04 & 0.128 & 0.391 \\
    A40.24   &  9 & 14 & 2.03 & 1.41 & 0.058 & 0.163 \\
    A60.24   &  8 &  9 & 0.51 & 0.91 & 0.769 & 0.487 \\
    A80.24   & 10 & 13 & 0.88 & 0.87 & 0.525 & 0.559 \\
    A100.24  & 10 & 10 & 0.84 & 0.94 & 0.554 & 0.473 \\
    \hline                                       
    A80.24s  & 11 & 12 & 0.56 & 0.86 & 0.815 & 0.561 \\
    A100.24s &  9 & 15 & 0.99 & 1.77 & 0.429 & 0.047 \\
    \hline                                       
    B25.32   & 13 &  8 & 0.94 & 0.84 & 0.495 & 0.518 \\
    B35.32   & 11 & 13 & 0.91 & 1.58 & 0.511 & 0.105 \\
    B55.32   &  8 &  8 & 0.69 & 0.83 & 0.628 & 0.529 \\
    B75.32   &  8 & 14 & 0.50 & 0.93 & 0.774 & 0.507 \\
    B85.24   &  9 &  8 & 0.86 & 1.27 & 0.521 & 0.272 \\
    \hline                                       
    D15.48   & 10 &  8 & 0.87 & 0.88 & 0.528 & 0.494 \\
    D20.48   &  9 &  8 & 0.88 & 0.64 & 0.506 & 0.666 \\
    D30.48   &  8 & 10 & 0.87 & 0.50 & 0.498 & 0.832 \\
    D45.32sc & 11 &  9 & 0.44 & 0.34 & 0.897 & 0.915 \\
    \hline\hline
    \vspace*{0.1cm}
  \end{tabular*}
  \caption{Final fit ranges to $\eta$ and $\eta'$ principal correlators together with resulting $\chi^2/\mathrm{dof}$ and $p$-values. All fits start at $t_1^{\eta, \eta'}=2a$; see text.}
  \label{tab:fit_params_final}
\end{table}

\begin{table}[t!]
 \centering
  \begin{tabular*}{1\textwidth}{@{\extracolsep{\fill}}lllll}
    \hline\hline
    ensemble  & $a\Getatogg \cdot 10^7$ & $a\Getaptogg \cdot 10^6$ & $a\hatFetagg$ & $a\hatFetapgg$ \\
    \hline\hline
    A30.32   & 3.65(45) & 5.5(1.6) & 0.015(09) & 0.1293(55) \\
    A40.32   & 2.90(23) & 2.4(0.2) & 0.053(03) & 0.1518(21) \\
    A40.24   & 2.63(19) & 2.3(0.2) & 0.047(03) & 0.1522(23) \\
    A60.24   & 2.62(18) & 2.6(0.4) & 0.045(06) & 0.1576(40) \\
    A80.24   & 2.24(16) & 3.1(0.4) & 0.041(04) & 0.1608(32) \\
    A100.24  & 1.83(14) & 2.3(0.2) & 0.051(04) & 0.1741(37) \\
    \hline                                                 
    A80.24s  & 1.23(17) & 2.5(0.6) & 0.040(06) & 0.1654(70) \\
    A100.24s & 1.16(11) & 2.8(0.3) & 0.034(03) & 0.1707(42) \\
    \hline                                                 
    B25.32   & 2.29(15) & 2.6(0.3) & 0.035(03) & 0.1285(17) \\
    B35.32   & 1.70(15) & 2.4(0.3) & 0.032(03) & 0.1376(24) \\
    B55.32   & 1.67(06) & 2.5(0.2) & 0.040(02) & 0.1457(16) \\
    B75.32   & 1.15(12) & 1.9(0.2) & 0.045(03) & 0.1583(28) \\
    B85.24   & 1.08(16) & 2.8(0.4) & 0.033(04) & 0.1507(39) \\
    \hline                                                 
    D15.48   & 2.38(25) & 1.4(0.5) & 0.044(08) & 0.0975(23) \\
    D20.48   & 2.26(28) & 2.1(0.8) & 0.046(09) & 0.1003(25) \\
    D30.48   & 2.21(34) & 1.7(0.9) & 0.042(12) & 0.1012(31) \\
    D45.32sc & 1.12(11) & 1.2(0.3) & 0.051(05) & 0.1109(32) \\
    \hline\hline
    \vspace*{0.1cm}
  \end{tabular*}
  \caption{Data for the decay widths computed from Eqs.~(\ref{eq:Geta2gg},\ref{eq:Getap2gg}) and the transition form factors in the large momentum limit as given in Eqs.~(\ref{eq:Fetagg},\ref{eq:Fetapgg}). Result are obtained using $Z$ factors from method M2, cf. Table~\ref{tab:beta_r0_a_Z}. Errors are statistical only.}
  \label{tab:lattice_results_P2gg_M2}
\end{table}

\end{document}